\pdfoutput=1
\documentclass[12pt,a4paper]{article}

\usepackage{ifthen} 
\newboolean{pdflatex}
\setboolean{pdflatex}{true} 

\newboolean{articletitles}
\setboolean{articletitles}{true} 

\newboolean{uprightparticles}
\setboolean{uprightparticles}{false} 

\def\paperauthors{LHCb collaboration} 
\def\paperasciititle{Measurement of the CKM angle gamma in B->DK and B->Dpi decays with D->KSh+h-} 
\def\papertitle{Measurement of the CKM angle $\gamma$\\ in $B^\pm\to D K^\pm$ and $B^\pm \to D \pi^\pm$ decays with $D \to K_\mathrm S^0 h^+ h^-$ } 
\def\paperkeywords{{High Energy Physics}, {LHCb}} 
\def\papercopyright{\the\year\ CERN for the benefit of the LHCb collaboration} 
\def\paperlicence{CC BY 4.0 licence}
\def\paperlicenceurl{https://creativecommons.org/licenses/by/4.0/}

\usepackage[top=1in, bottom=1.25in, left=1in, right=1in]{geometry}

\usepackage{microtype}
\usepackage{lineno}  
\usepackage{xspace} 
\usepackage{caption} 

\usepackage{graphicx}  
\usepackage{color}
\usepackage{colortbl}
\graphicspath{{./figs/}} 

\usepackage{amsmath} 
\usepackage{amssymb}
\usepackage{amsfonts}
\usepackage{upgreek} 

\newcommand*\patchAmsMathEnvironmentForLineno[1]{%
\expandafter\let\csname old#1\expandafter\endcsname\csname #1\endcsname
\expandafter\let\csname oldend#1\expandafter\endcsname\csname
end#1\endcsname
 \renewenvironment{#1}%
   {\linenomath\csname old#1\endcsname}%
   {\csname oldend#1\endcsname\endlinenomath}%
}
\newcommand*\patchBothAmsMathEnvironmentsForLineno[1]{%
  \patchAmsMathEnvironmentForLineno{#1}%
  \patchAmsMathEnvironmentForLineno{#1*}%
}
\AtBeginDocument{%
\patchBothAmsMathEnvironmentsForLineno{equation}%
\patchBothAmsMathEnvironmentsForLineno{align}%
\patchBothAmsMathEnvironmentsForLineno{flalign}%
\patchBothAmsMathEnvironmentsForLineno{alignat}%
\patchBothAmsMathEnvironmentsForLineno{gather}%
\patchBothAmsMathEnvironmentsForLineno{multline}%
\patchBothAmsMathEnvironmentsForLineno{eqnarray}%
}

\usepackage{hyperxmp}
\usepackage{booktabs}

\usepackage[pdftex,
            pdfauthor={\paperauthors},
            pdftitle={\paperasciititle},
            pdfkeywords={\paperkeywords},
            pdfcopyright={Copyright (C) \papercopyright},
            pdflicenseurl={\paperlicenceurl}]{hyperref}

\usepackage[colorinlistoftodos,textsize=scriptsize]{todonotes}

\usepackage[bottom,flushmargin,hang,multiple]{footmisc}

\usepackage[all]{hypcap} 

\usepackage{xspace} 
\usepackage{upgreek}

\def\lhcb   {\mbox{LHCb}\xspace}

\def\babar  {\mbox{BaBar}\xspace}
\def\belle  {\mbox{Belle}\xspace}

\def\besiii {\mbox{BESIII}\xspace}
\def\cleo   {\mbox{CLEO}\xspace}

\def\MagUp {\mbox{\em Mag\kern -0.05em Up}\xspace}

\ifthenelse{\boolean{uprightparticles}}%
{
 \def\Pbeta       {\ensuremath{\upbeta}\xspace}
 \def\Pgamma      {\ensuremath{\upgamma}\xspace}

 \def\Ppi         {\ensuremath{\uppi}\xspace}

 \def\Ppsi        {\ensuremath{\uppsi}\xspace}

 \def\PDelta      {\ensuremath{\Delta}\xspace}                 
 \def\PXi         {\ensuremath{\Xi}\xspace}                 
 \def\PLambda     {\ensuremath{\Lambda}\xspace}                 
 \def\PSigma      {\ensuremath{\Sigma}\xspace}                 
 \def\POmega      {\ensuremath{\Omega}\xspace}                 
 \def\PUpsilon    {\ensuremath{\Upsilon}\xspace}

 \def\PB      {\ensuremath{\mathrm{B}}\xspace}                 
                  
 \def\PD      {\ensuremath{\mathrm{D}}\xspace}

 \def\PK      {\ensuremath{\mathrm{K}}\xspace}

 \def\Pb      {\ensuremath{\mathrm{b}}\xspace}                 
 \def\Pc      {\ensuremath{\mathrm{c}}\xspace}                 
 \def\Pd      {\ensuremath{\mathrm{d}}\xspace}

 \def\Ph      {\ensuremath{\mathrm{h}}\xspace}                 
 \def\Pi      {\ensuremath{\mathrm{i}}\xspace}

 \def\Pp      {\ensuremath{\mathrm{p}}\xspace}

 \def\Ps      {\ensuremath{\mathrm{s}}\xspace}                 
                  
 \def\Pu      {\ensuremath{\mathrm{u}}\xspace}

 \def\thebaroffset{0.0em}
}
{
 \def\Pbeta       {\ensuremath{\beta}\xspace}
 \def\Pgamma      {\ensuremath{\gamma}\xspace}

 \def\Ppi         {\ensuremath{\pi}\xspace}

 \def\Ppsi        {\ensuremath{\psi}\xspace}                 
                  
 \mathchardef\PDelta="7101
 \mathchardef\PXi="7104
 \mathchardef\PLambda="7103
 \mathchardef\PSigma="7106
 \mathchardef\POmega="710A
 \mathchardef\PUpsilon="7107
                  
 \def\PB      {\ensuremath{B}\xspace}                 
                  
 \def\PD      {\ensuremath{D}\xspace}

 \def\PK      {\ensuremath{K}\xspace}

 \def\Pb      {\ensuremath{b}\xspace}                 
 \def\Pc      {\ensuremath{c}\xspace}                 
 \def\Pd      {\ensuremath{d}\xspace}

 \def\Ph      {\ensuremath{h}\xspace}                 
 \def\Pi      {\ensuremath{i}\xspace}

 \def\Pp      {\ensuremath{p}\xspace}

 \def\Ps      {\ensuremath{s}\xspace}                 
                  
 \def\Pu      {\ensuremath{u}\xspace}

 \def\thebaroffset{0.18em}
}
\newcommand{\offsetoverline}[2][\thebaroffset]{\kern #1\overline{\kern -#1 #2}}%

\makeatletter
\ifcase \@ptsize \relax
  \newcommand{\miniscule}{\@setfontsize\miniscule{4}{5}}
\or
  \newcommand{\miniscule}{\@setfontsize\miniscule{5}{6}}
\or
  \newcommand{\miniscule}{\@setfontsize\miniscule{5}{6}}
\fi
\makeatother

\DeclareRobustCommand{\optbar}[1]{\shortstack{{\miniscule (\rule[.5ex]{1.25em}{.18mm})}
  \\ [-.7ex] $#1$}}

\def\g      {{\ensuremath{\Pgamma}}\xspace}

\def\uquark    {{\ensuremath{\Pu}}\xspace}
\def\uquarkbar {{\ensuremath{\overline \uquark}}\xspace}

\def\dquark    {{\ensuremath{\Pd}}\xspace}
\def\dquarkbar {{\ensuremath{\overline \dquark}}\xspace}

\def\squark    {{\ensuremath{\Ps}}\xspace}
\def\squarkbar {{\ensuremath{\overline \squark}}\xspace}

\def\cquark    {{\ensuremath{\Pc}}\xspace}
\def\cquarkbar {{\ensuremath{\overline \cquark}}\xspace}

\def\bquark    {{\ensuremath{\Pb}}\xspace}
\def\bquarkbar {{\ensuremath{\overline \bquark}}\xspace}

\def\hadron {{\ensuremath{\Ph}}\xspace}
\def\pion   {{\ensuremath{\Ppi}}\xspace}
\def\piz    {{\ensuremath{\pion^0}}\xspace}
\def\pip    {{\ensuremath{\pion^+}}\xspace}
\def\pim    {{\ensuremath{\pion^-}}\xspace}
\def\pipm   {{\ensuremath{\pion^\pm}}\xspace}
\def\pimp   {{\ensuremath{\pion^\mp}}\xspace}

\def\kaon    {{\ensuremath{\PK}}\xspace}

\def\KorKbar {\kern \thebaroffset\optbar{\kern -\thebaroffset \PK}{}\xspace}

\def\Kp      {{\ensuremath{\kaon^+}}\xspace}
\def\Km      {{\ensuremath{\kaon^-}}\xspace}
\def\Kpm     {{\ensuremath{\kaon^\pm}}\xspace}

\def\KS      {{\ensuremath{\kaon^0_{\mathrm{S}}}}\xspace}

\def\Dbar    {{\ensuremath{\offsetoverline{\PD}}}\xspace}
\def\D       {{\ensuremath{\PD}}\xspace}

\def\DorDbar {\kern \thebaroffset\optbar{\kern -\thebaroffset \PD}\xspace}
\def\Dz      {{\ensuremath{\D^0}}\xspace}
\def\Dzb     {{\ensuremath{\Dbar{}^0}}\xspace}
\def\Dp      {{\ensuremath{\D^+}}\xspace}
\def\Dm      {{\ensuremath{\D^-}}\xspace}

\def\DpDm    {\ensuremath{\Dp {\kern -0.16em \Dm}}\xspace}

\def\B       {{\ensuremath{\PB}}\xspace}

\def\BorBbar {\kern \thebaroffset\optbar{\kern -\thebaroffset \PB}\xspace}
\def\Bz      {{\ensuremath{\B^0}}\xspace}

\def\Bd      {{\ensuremath{\B^0}}\xspace}

\def\BdorBdbar {\kern \thebaroffset\optbar{\kern -\thebaroffset \Bd}\xspace}
\def\Bu      {{\ensuremath{\B^+}}\xspace}
\def\Bub     {{\ensuremath{\B^-}}\xspace}
\def\Bp      {{\ensuremath{\Bu}}\xspace}
\def\Bm      {{\ensuremath{\Bub}}\xspace}
\def\Bpm     {{\ensuremath{\B^\pm}}\xspace}

\def\Bs      {{\ensuremath{\B^0_\squark}}\xspace}

\def\BsorBsbar {\kern \thebaroffset\optbar{\kern -\thebaroffset \Bs}\xspace}

\def\psiprpr  {{\ensuremath{\Ppsi(3770)}}\xspace}

\def\Y#1S{\ensuremath{\PUpsilon{(#1S)}}\xspace}

\def\Lz          {{\ensuremath{\PLambda}}\xspace}

\def\LorLbar     {\kern \thebaroffset\optbar{\kern -\thebaroffset \PLambda}\xspace}

\def\Lc          {{\ensuremath{\Lz^+_\cquark}}\xspace}

\def\Lb           {{\ensuremath{\Lz^0_\bquark}}\xspace}

\newcommand{\decay}[2]{\ensuremath{#1\!\to #2}\xspace} 

\def\to                 {\ensuremath{\rightarrow}\xspace}

\def\CP                {{\ensuremath{C\!P}}\xspace}

\def\Vud  {{\ensuremath{V_{\uquark\dquark}^{\phantom{\ast}}}}\xspace}
\def\Vcd  {{\ensuremath{V_{\cquark\dquark}^{\phantom{\ast}}}}\xspace}

\def\Vubs  {{\ensuremath{V_{\uquark\bquark}^\ast}}\xspace}
\def\Vcbs  {{\ensuremath{V_{\cquark\bquark}^\ast}}\xspace}

\def\AT#1     {\ensuremath{A_{\mathrm{T}}^{#1}}\xspace}           

\def\C#1      {\ensuremath{\mathcal{C}_{#1}}\xspace}                       
\def\Cp#1     {\ensuremath{\mathcal{C}_{#1}^{'}}\xspace}                    
\def\Ceff#1   {\ensuremath{\mathcal{C}_{#1}^{\mathrm{(eff)}}}\xspace}        
\def\Cpeff#1  {\ensuremath{\mathcal{C}_{#1}^{'\mathrm{(eff)}}}\xspace}       
\def\Ope#1    {\ensuremath{\mathcal{O}_{#1}}\xspace}                       
\def\Opep#1   {\ensuremath{\mathcal{O}_{#1}^{'}}\xspace}                    

\newcommand{\nospaceunit}[1]{\ensuremath{\text{#1}}}       
\newcommand{\aunit}[1]{\ensuremath{\text{\,#1}}}       

\newcommand{\tev}{\aunit{Te\kern -0.1em V}\xspace}
\newcommand{\gev}{\aunit{Ge\kern -0.1em V}\xspace}
\newcommand{\mev}{\aunit{Me\kern -0.1em V}\xspace}
\newcommand{\kev}{\aunit{ke\kern -0.1em V}\xspace}
\newcommand{\ev}{\aunit{e\kern -0.1em V}\xspace}
 
\newcommand{\mevc}{\ensuremath{\aunit{Me\kern -0.1em V\!/}c}\xspace}
\newcommand{\gevc}{\ensuremath{\aunit{Ge\kern -0.1em V\!/}c}\xspace}
\newcommand{\mevcc}{\ensuremath{\aunit{Me\kern -0.1em V\!/}c^2}\xspace}
\newcommand{\gevcc}{\ensuremath{\aunit{Ge\kern -0.1em V\!/}c^2}\xspace}

\def\mum  {\ensuremath{\,\upmu\nospaceunit{m}}\xspace}

\def\fb   {\ensuremath{\aunit{fb}}\xspace}
\def\invfb   {\ensuremath{\fb^{-1}}\xspace}

\def\ci {\aunit{Ci}\xspace}

\newcommand{\chisq}{\ensuremath{\chi^2}\xspace}

\newcommand{\chisqip}{\ensuremath{\chi^2_{\text{IP}}}\xspace}

\def\gsim{{~\raise.15em\hbox{$>$}\kern-.85em
          \lower.35em\hbox{$\sim$}~}\xspace}
\def\lsim{{~\raise.15em\hbox{$<$}\kern-.85em
          \lower.35em\hbox{$\sim$}~}\xspace}

\def\sqs   {\ensuremath{\protect\sqrt{s}}\xspace}

\def\pt         {\ensuremath{p_{\mathrm{T}}}\xspace}

\def\ptot       {\ensuremath{p}\xspace}

\def\evtgen     {\mbox{\textsc{EvtGen}}\xspace}

\def\geant      {\mbox{\textsc{Geant4}}\xspace}

\def\photos     {\mbox{\textsc{Photos}}\xspace}

\def\pythia     {\mbox{\textsc{Pythia}}\xspace}

\def\tell1  {TELL1\xspace}
\def\ukl1   {UKL1\xspace}

\newcommand{\Kspipi}{\ensuremath{\KS\pip\pim}\xspace}
\newcommand{\KsPiPi}{\ensuremath{\KS\pip\pim}\xspace}
\newcommand{\KsKK}{\ensuremath{\KS\Kp\Km}\xspace}
\newcommand{\Kshh}{\ensuremath{\KS\Ph^+\Ph^-}\xspace}

\newcommand{\DtoKspipi}{\ensuremath{\PD\to\Kspipi}\xspace}
\newcommand{\DtoKsKK}{\ensuremath{\PD\to\KsKK}\xspace}
\newcommand{\DtoKshh}{\ensuremath{\PD\to\Kshh}\xspace}

\newcommand{\BtoDpi}{\ensuremath{\Bpm\to\PD\pipm}\xspace}
\newcommand{\BtoDK}{\ensuremath{\Bpm\to\PD\Kpm}\xspace}
\newcommand{\BtoDh}{\ensuremath{\PB\to\PD\Ph}\xspace}

\newcommand{\Dpi}{\ensuremath{\PD\pi^\pm}\xspace}

\newcommand{\DK}{\ensuremath{\PD\PK^\pm}\xspace}

\newcommand{\Ks}{\ensuremath{\KS}\xspace}
\newcommand{\msqmin}{\ensuremath{m^2_-}\xspace}
\newcommand{\msqplus}{\ensuremath{m^2_+}\xspace}

\def        \BtoDpi         {\mbox{\ensuremath{{\Bpm\to\D\pipm}}}\xspace}
\def        \BtoDK          {\mbox{\ensuremath{{\Bpm\to\D\Kpm}}}\xspace}

\def        \DK             {{\D\Kpm}\xspace}
\def        \BtoDh          {\ensuremath{\Bpm\to\D\hadron^\pm}\xspace}
\def        \DtoKspp        {\mbox{\ensuremath{\D\to\KS\pip\pim}}\xspace}
\def        \DtoKskk        {\mbox{\ensuremath{\D\to\KS\Kp\Km}}\xspace}
\def        \Kshh           {\ensuremath{\D\to\KS\hadron^+\hadron^-}\xspace}
\def        \DtoKshh        {\ensuremath{\D\to\KS\hadron^+\hadron^-}\xspace}

\newcommand{\xpm}{\ensuremath{x_\pm}\xspace}
\newcommand{\ypm}{\ensuremath{y_\pm}\xspace}
\newcommand{\xm}{\ensuremath{x_-}\xspace}
\newcommand{\ym}{\ensuremath{y_-}\xspace}
\newcommand{\xp}{\ensuremath{x_+}\xspace}
\newcommand{\yp}{\ensuremath{y_+}\xspace}

\newcommand{\xpmdk}{\ensuremath{x_\pm^{\D\kaon}}\xspace}
\newcommand{\ypmdk}{\ensuremath{y_\pm^{\D\kaon}}\xspace}
\newcommand{\xpmdpi}{\ensuremath{x_\pm^{\D\pi}}\xspace}
\newcommand{\ypmdpi}{\ensuremath{y_\pm^{\D\pi}}\xspace}
\newcommand{\xmdk}{\ensuremath{x_-^{\D\kaon}}\xspace}
\newcommand{\ymdk}{\ensuremath{y_-^{\D\kaon}}\xspace}
\newcommand{\xpdk}{\ensuremath{x_+^{\D\kaon}}\xspace}
\newcommand{\ypdk}{\ensuremath{y_+^{\D\kaon}}\xspace}
\newcommand{\xxidpi}{\ensuremath{x_\xi^{\D\pi}}\xspace}
\newcommand{\yxidpi}{\ensuremath{y_\xi^{\D\pi}}\xspace}

\renewcommand{\Re}{\text{Re}}
\renewcommand{\Im}{\text{Im}}

\newcommand{\dpi}{\ensuremath{\D\pi}\xspace}

\renewcommand{\g}{\ensuremath{\gamma}\xspace}
\newcommand{\rB}{\ensuremath{r_\B}\xspace}

\newcommand{\rBDK}{\ensuremath{r_\B^{\D \kaon}}\xspace}
\newcommand{\rBDpi}{\ensuremath{r_\B^{\D\pi}}\xspace}

\newcommand{\dB}{\ensuremath{\delta_\B}\xspace}
\newcommand{\dBDK}{\ensuremath{\delta_\B^{\D\kaon}}\xspace}
\newcommand{\dBDpi}{\ensuremath{\delta_\B^{\D\pi}}\xspace}

\renewcommand{\ci}{\ensuremath{c_i}\xspace}
\newcommand{\si}{\ensuremath{s_i}\xspace}

\newcommand{\Fi}{\ensuremath{F_i}\xspace}

\usepackage{cite} 
\usepackage{mciteplus}

\usepackage{longtable} 

\begin{document}

\renewcommand{\thefootnote}{\fnsymbol{footnote}}
\setcounter{footnote}{1}


\begin{titlepage}
\pagenumbering{roman}

\vspace*{-1.5cm}
\centerline{\large EUROPEAN ORGANIZATION FOR NUCLEAR RESEARCH (CERN)}
\vspace*{1.5cm}
\noindent
\begin{tabular*}{\linewidth}{lc@{\extracolsep{\fill}}r@{\extracolsep{0pt}}}
\ifthenelse{\boolean{pdflatex}}
{\vspace*{-1.5cm}\mbox{\!\!\!\includegraphics[width=.14\textwidth]{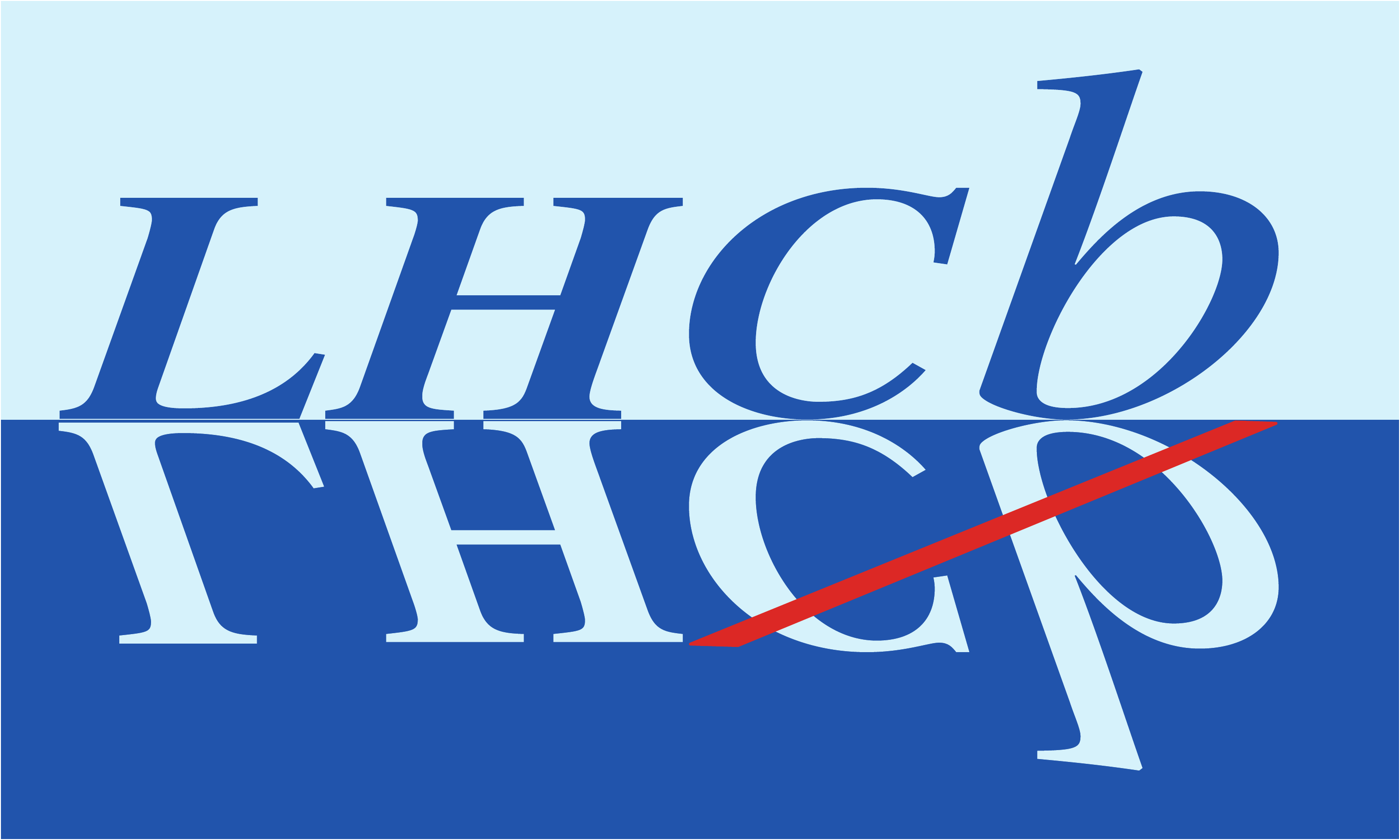}} & &}%
{\vspace*{-1.2cm}\mbox{\!\!\!\includegraphics[width=.12\textwidth]{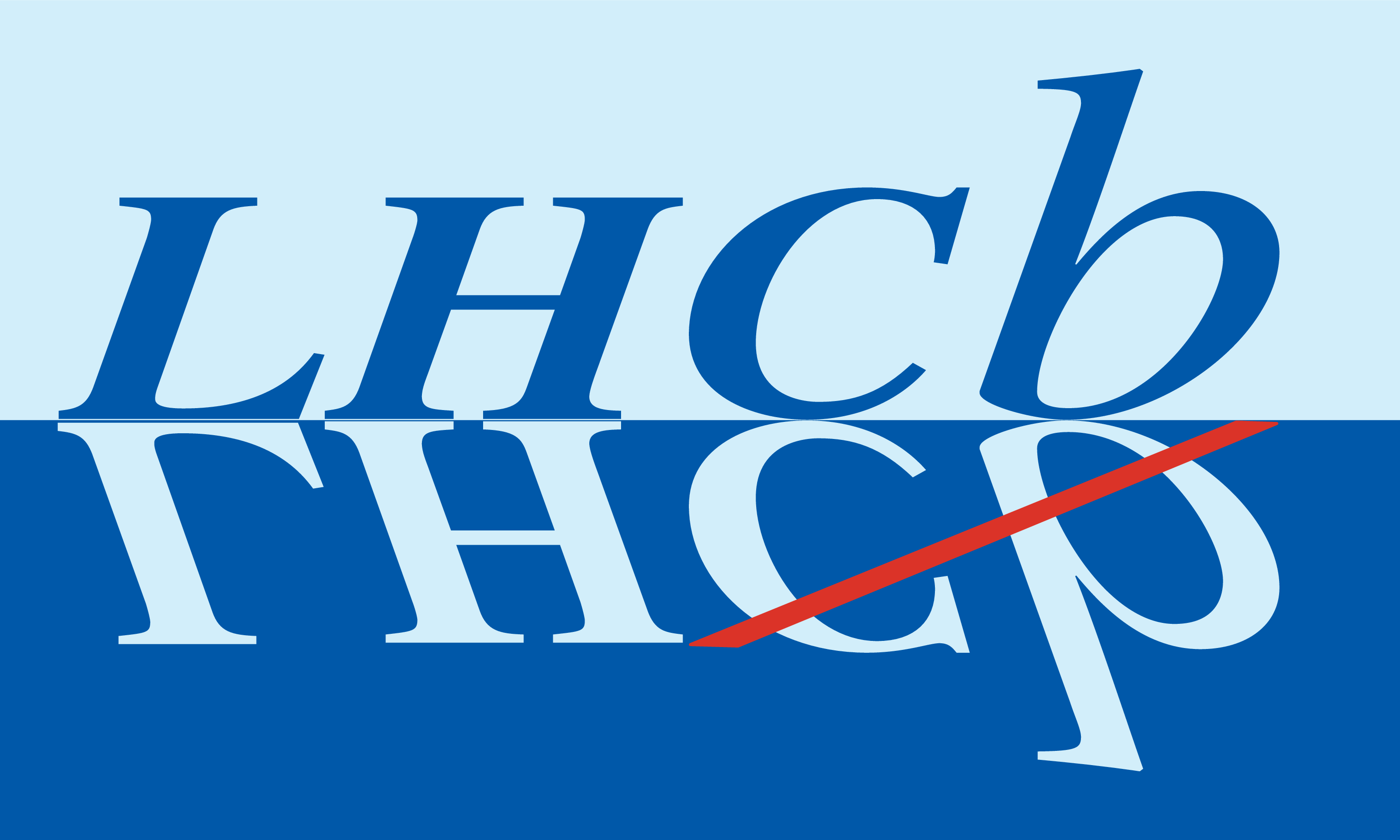}} & &}%
\\
 & & CERN-EP-2020-175 \\  
 & & LHCb-PAPER-2020-019 \\  
 & & \today \\ 
 & & \\
\end{tabular*}

\vspace*{4cm}

{\normalfont\bfseries\boldmath\huge
\begin{center}
  \papertitle 
\end{center}
}

\vspace*{2.0cm}

\begin{center}
\paperauthors\footnote{Authors are listed at the end of this paper.}
\end{center}


\vspace{\fill}
\begin{abstract}
  \noindent
   A measurement of \CP-violating observables is performed using the decays \BtoDK and \BtoDpi, where the \D meson is reconstructed in one of the self-conjugate three-body final states \KsPiPi and \KsKK (commonly denoted $\KS h^+h^-$). The decays are analysed in bins of the \D-decay phase space, leading to a measurement that is independent of the modelling of the \D-decay amplitude. The observables are interpreted in terms of the CKM angle $\gamma$. Using a data sample corresponding to an integrated luminosity of 9\invfb collected in proton-proton collisions at centre-of mass energies of $7$, $8$, and $13\tev$ with the \lhcb experiment, \Pgamma is measured to be $\left(68.7^{+5.2}_{-5.1}\right)^\circ$. The hadronic parameters \rBDK, \rBDpi, \dBDK, and \dBDpi, which are the ratios and strong-phase differences of the suppressed and favoured \Bpm decays, are also reported.
   

\end{abstract}

\vspace*{2.0cm}

\begin{center}
  Published in 
  J. High Energ. Phys. 2021, 169 (2021).
\end{center}

\vspace{\fill}

{\footnotesize 
\centerline{\copyright~\papercopyright. \href{\paperlicenceurl}{\paperlicence}.}}
\vspace*{2mm}

\end{titlepage}


\newpage
\setcounter{page}{2}
\mbox{~}
%
%
%
%


\renewcommand{\thefootnote}{\arabic{footnote}}
\setcounter{footnote}{0}

\cleardoublepage


\pagestyle{plain} 
\setcounter{page}{1}
\pagenumbering{arabic}



\section{Introduction}
\label{sec:intro}

In the framework of the Standard Model, \CP violation can be described by the angles and lengths of the Unitarity Triangle constructed from elements of the CKM matrix~\cite{Cabibbo:1963yz,Kobayashi:1973fv}. The angle $\g \equiv \textrm{arg}\left(-\Vud\Vubs/\Vcd\Vcbs\right)$ has particularly interesting features. It is the only CKM angle that can be measured in decays including only tree-level processes, and is experimentally accessible through the interference of $\bquarkbar \to \cquarkbar\uquark\squarkbar$ and $\bquarkbar \to \uquarkbar\cquark\squarkbar$ (and \CP-conjugate) decay amplitudes. In addition, there are negligible theoretical uncertainties when interpreting the measured observables in terms of \g~\cite{BrodZup}. Hence, in the absence of unknown physics effects at tree level, a precision measurement of \g provides a Standard Model benchmark that can be compared with indirect determinations from other CKM-matrix observables more likely to be affected by physics beyond the Standard Model~\cite{Blanke:2018cya}. Such comparisons are currently limited by the precision of direct measurements of \g, which is about $5^\circ$~\cite{HFLAV18,CKMfitter2005} dominated by \lhcb results.

Decays such as \BtoDK, where \D represents a superposition of \Dz and \Dzb states, are used to observe the effects of interference between $\bquarkbar \to \cquarkbar\uquark\squarkbar$ and $\bquarkbar \to \uquarkbar\cquark\squarkbar$ (and \CP-conjugate) decay amplitudes. The interference arises when the decay channel of the \D meson is common to both \Dz and \Dzb mesons. The \BtoDK decay has been studied extensively with a wide range of \D-meson final states~\cite{LHCb-PAPER-2019-044,LHCb-PAPER-2016-003,LHCb-PAPER-2015-014,LHCb-PAPER-2018-017, BelleKspipipi0}. The exact choice of observables from each of these analyses is dependent on the method that is most appropriate for the \D decay used~\cite{ads,ads2,GLW1,GLW2,qGLW,GGSZ,BONDARGGSZ,AGGSZ,GLS}. The methods can be extended to a variety of different \B-decay modes~\cite{LHCb-PAPER-2016-003,LHCb-PAPER-2016-006,LHCb-PAPER-2017-030,LHCb-PAPER-2015-020, LHCb-PAPER-2017-021}.

This paper presents a model-independent study of the decay modes \BtoDK and \BtoDpi where the chosen \D decays are the self-conjugate decays \DtoKspp and \DtoKskk (denoted \DtoKshh). The analysis of the \BtoDK, \DtoKshh decay chain is powerful due to the rich resonance structure of the \D-decay modes, as has been described in Refs.~\cite{BONDARGGSZ,GGSZ,AGGSZ}. 
The data used in this analysis were accumulated with the \lhcb detector over the period 2011--2018 in $pp$ collisions at energies of \sqs=7,\,8,\,13 \tev, corresponding to a total integrated luminosity of approximately 9\invfb.

The presence of interference leads to differences in the phase-space distributions of
\D decays from reconstructed \Bp and \Bm decays. In order to interpret any observed difference in the context of the angle \g, knowledge of the strong phase of the \Dz decay amplitude, and how it varies over phase space, is required. An attractive model-independent approach makes use of direct measurements of the strong-phase difference between \Dz and \Dzb decays, averaged over regions of the phase space~\cite{GGSZ,BPMODIND1,BPMODIND2}. Quantum correlated pairs of \D mesons produced in decays of \psiprpr give direct access to the strong-phase differences. These have been measured by the \cleo collaboration~\cite{CLEOCISI}, and more recently the \besiii collaboration~\cite{bes3prl,bes3prd,Krishna}. Measurements using the inputs in Ref.~\cite{CLEOCISI} have been used by the \lhcb~\cite{LHCb-PAPER-2018-017,LHCb-PAPER-2016-006,LHCb-PAPER-2014-041} and \belle~\cite{BELLEMODIND,BelDKst} collaborations. An alternate method is to use an amplitude model of the \D decay to determine the strong-phase variation~\cite{delAmoSanchez:2010rq,Poluektov:2010wz,LHCb-PAPER-2014-017}. 
The separation of data into binned regions of the Dalitz plot leads to a loss of statistical sensitivity in comparison to using an amplitude model. However, the advantage of using the direct strong-phase measurements resides in the model-independent nature of the systematic uncertainties. Where the direct strong-phase measurements are used, there is only a systematic uncertainty associated with the finite precision of such measurements. Conversely, systematic uncertainties associated with determining a phase from an amplitude model are difficult to evaluate, as common approaches to amplitude-model building violate the optical theorem~\cite{Battaglieri:2014gca}. Therefore, the loss in statistical precision is compensated by reliability in the evaluation of the systematic uncertainty, which is increasingly important as the overall precision on the CKM angle \g improves.
The analysis approach is laid out in Sect.~\ref{sec:theory}, while Sect.~\ref{sec:detector} describes the \lhcb detector used to collect the data sample, and Sect.~\ref{sec:selection} summarises the  selection criteria. The measurement is based on a two-step fit procedure covered in Sect.~\ref{sec:massfit}, where the fit to the invariant-mass distribution is detailed, and Sect.~\ref{sec:cpfit}, which describes how the \CP observables are determined. The systematic uncertainties are reported in Sect.~\ref{sec:syst}, and the results are interpreted to determine the value of $\gamma$ in Sect.~\ref{sec:interpretation}. Finally, the conclusions are presented in Sect.~\ref{sec:conclusions}.

\section{Analysis Overview}
\label{sec:theory}
The sum of the favoured and suppressed contributions to the $\Bm \to\D\Km$ amplitude can be written as
\begin{equation}
A_B (\msqmin,\msqplus) \propto \, A_D(\msqmin,\msqplus) + \rBDK e^{i(\dBDK - \g)}A_{\Dbar}(\msqmin,\msqplus) ,
\label{eq:bamplitude}
\end{equation}
where  $A_D(\msqmin,\msqplus)$ is the $\Dz \to \KS h^+ h^-$ decay amplitude, and $A_{\Dbar}(\msqmin,\msqplus)$ is the  ${\Dzb \to \KS h^+ h^-}$ decay amplitude.  The hadronic parameters \rBDK and \dBDK are the ratio of the magnitudes of the amplitudes of $\Bm \to \Dzb \Km$ and $\Bm \to \Dz \Km$ and the strong-phase difference between them, respectively. Finally, the position of the decay in the Dalitz plot is defined by \msqmin and \msqplus, which are the squared invariant masses of the $\KS h^-$ and  $\KS h^+$ particle combinations, respectively. The equivalent expression for the charge-conjugated decay $B^+ \to D K^+$ is obtained by making the substitutions $\gamma \to -\gamma$ and $A_D(\msqmin,\msqplus) \leftrightarrow A_{\Dbar}(\msqmin,\msqplus)$.

The $D$-decay phase space is partitioned into $2\times \mathcal{N}$ bins labelled from $i=-\mathcal{N}$ to $i=+\mathcal{N}$ (excluding zero), symmetric around  $\msqmin=\msqplus$  such that if $(\msqmin ,\msqplus)$ is in bin $i$ then $(\msqplus,\msqmin)$ is in bin $-i$. The bins for which $\msqmin > \msqplus$ are defined to have positive values of $i$\footnote{For historical reasons, this convention defines positive bins in the opposite manner to that used to determine the charm strong-phase differences in \DtoKshh decays. }.
The strong-phase difference between the \Dz- and \Dzb-decay amplitudes at a given point on the Dalitz plot is denoted as $\delta_D(m_-^2,m_+^2)$.
The cosine of $\delta_D(m_-^2,m_+^2)$ weighted by the \D-decay amplitude and averaged
over bin $i$ is written as $c_i$~\cite{GGSZ}, and is given by
\begin{align}
c_i &\equiv \frac{\int_{i} d\msqmin \, d\msqplus \, |A_D(\msqmin,\msqplus)| |A_D(\msqplus,\msqmin)| \cos\
[\delta_D(\msqmin,\msqplus)-\delta_D(\msqplus,\msqmin)]}
{\sqrt{\int_{i} d\msqmin \, d\msqplus \, |A_D(\msqmin,\msqplus)|^2 \int_{i} d\msqmin \, d\msqplus \, |A_D(\msqplus,\msqmin)|^2}}\,,
\label{eq:ci}
\end{align}
where the integrals are evaluated over bin $i$. An analogous expression can be written for $s_i$, which is the sine of the strong-phase difference weighted by the decay amplitude and averaged over the bin phase space.

The expected yield of \Bm decays in bin $i$ is found by integrating the square of the amplitude given in  Eq.~\eqref{eq:bamplitude} over the region of phase space defined by the $i$th bin. The effects of charm mixing and \CP violation are ignored, as is the presence of \CP violation and matter regeneration in the neutral $K^0$ decays. These effects are expected to have a small impact~\cite{BPV,KsCPV} on the distribution of events on the Dalitz plot. Selection requirements and reconstruction effects lead to a non-uniform efficiency over  phase space, denoted by $\eta(\msqmin,\msqplus)$. At \lhcb the typical efficiency variation over phase space for a \DtoKshh decay from a region of high efficiency to low efficiency is approximately 60\%~\cite{LHCb-PAPER-2016-006}. The fractional yield of pure \Dz decays in bin $i$ in the presence of this efficiency profile is denoted \Fi, given by
\begin{equation}
\label{eq:fi}
\Fi = \frac{\int_{i} d\msqmin d\msqplus |A_{D}(\msqmin,\msqplus)|^2 \, \eta(\msqmin,\msqplus) }{\sum_j \int_{j} d\msqmin d\msqplus |A_{D}(\msqmin,\msqplus)|^2\, \eta(\msqmin,\msqplus) },
\end{equation}
where the sum in the denominator is over all Dalitz plot bins, indexed by $j$. Neglecting \CP violation in these charm decays, the charge-conjugate amplitudes satisfy the relation $A_{\Dbar}(\msqmin,\msqplus) = A_D(\msqplus,\msqmin)$, and therefore \Fi = $\overline{F}_{-i}$, where $\overline{F}_i$ is the fractional yield of \Dzb decays to bin $i$. The physics parameters of interest, $\rBDK$, $\dBDK$, and $\g$, are translated into four \CP-violating observables~\cite{BABAR2005} that are measured in this analysis and are the real and imaginary parts of the ratio of the suppressed and favoured \B decay amplitudes,
\begin{equation}
\xpmdk \equiv \rBDK \cos (\dBDK \pm \gamma) {\rm\ \ and\ } \; \ypmdk \equiv \rBDK \sin (\dBDK \pm \gamma).
\label{eq:xydefinitions}
\end{equation}
Using the relations $\ci = c_{-i}$ and $\si= - s_{-i}$ the \Bp (\Bm) yields, $N^+$ ($N^-$), in bin $i$ and $-i$ are given by
\begin{align}
\begin{split}
N_{+i}^+ &= h_{B^+} \left[ F_{-i} + \left(\left(\xpdk\right)^2 + \left(\ypdk\right)^2\right) F_{+ i} + 2 \sqrt{F_i F_{-i}} \left( \xpdk c_{+i} - \ypdk s_{+i}\right) \right], \\
N_{-i}^+ &= h_{B^+} \left[ F_{+i} + \left(\left(\xpdk\right)^2 + \left(\ypdk\right)^2\right) F_{- i} + 2 \sqrt{F_i F_{-i}} \left( \xpdk c_{+i} + \ypdk s_{+i}\right) \right], \\
N_{+i}^- &= h_{B^-} \left[ F_{+i} + \left(\left(\xmdk\right)^2 + \left(\ymdk\right)^2\right) F_{- i} + 2 \sqrt{F_i F_{-i}} \left( \xmdk c_{+i} + \ymdk s_{+i}\right) \right], \\
N_{-i}^- &= h_{B^-} \left[ F_{-i} + \left(\left(\xmdk\right)^2 + \left(\ymdk\right)^2\right) F_{+ i} + 2 \sqrt{F_i F_{-i}} \left( \xmdk c_{+i} - \ymdk s_{+i}\right) \right], 
\label{eq:populations}
\end{split}
\end{align}
where $h_{B^+}$ and $h_{B^-}$ are normalisation constants. The value of $\rBDK$ is allowed to be different for each charge and is constructed from either $(\rBDK)^2 = \left(\xpdk\right)^2 + \left(\ypdk\right)^2$ or $(\rBDK)^2 = \left(\xmdk\right)^2 + \left(\ymdk\right)^2$. A single value of \rBDK is determined when the \CP observables are subsequently interpreted to determine the physics parameters of interest. The normalisation constants can be written as a function of \Pgamma, analogous to the global asymmetries studied in decays where the \D meson decays to a \CP eigenstate~\cite{LHCb-PAPER-2016-003}. However, not only is this global asymmetry expected to be small since the \CP-even content of the \DtoKspp and \DtoKskk decay modes is close to 0.5, it is also expected to be heavily biased due to the effects of $\KS$ \CP violation~\cite{KsCPV} on total yields. Therefore the global asymmetry is ignored and the loss of information is minimal. An advantage of this approach is that the normalisation constants $h_{\Bp}$ and $h_{\Bm}$ are independent of each other, and will implicitly contain the effects of the production asymmetry of \Bpm mesons in $pp$ collisions and the detection asymmetries of the charged kaon from the \B decay. This leads to a \CP-violation measurement that is free of systematic uncertainties associated to these effects. 

The system of equations provides $4\mathcal{N}$ observables and $4+2\mathcal{N}$ unknowns, assuming that the available measurements of \ci and \si are used. This is solvable for $\mathcal{N}\geq2$, but in practice the simultaneous fit of the \Fi, \xpmdk, and \ypmdk parameters leads to large uncertainties on the \CP observables, and hence some external knowledge of the \Fi parameters is desirable. The \Fi parameters could be computed from simulation and an amplitude model, but the systematic uncertainties associated with the \lhcb simulation would be significant. Recent analyses~\cite{LHCb-PAPER-2014-041,LHCb-PAPER-2018-017} have used the semileptonic decay $B\to \D^*\mu\nu$, where the flavour-tagged yields of \Dz mesons are corrected for the differences in selection between the semileptonic channel and the signal mode. However, with the increased signal yields, the uncertainty due to this necessary correction will be approximately half the statistical uncertainty on the measurement presented in this paper, and therefore a different  method is adopted. 

The \BtoDpi decay mode is expected to have \Fi parameters that are the same as those for \BtoDK if a similar selection is applied due to the common topology and the ability to use same signatures in the detector to select the candidates. The \BtoDpi decay is expected to exhibit \CP violation through the interference of $\bquarkbar \to \cquarkbar\uquark\dquarkbar$ and $\bquarkbar \to \uquarkbar\cquark\dquarkbar$ transitions, analogous to the \BtoDK decay but suppressed by one order of magnitude~\cite{Kenzie:2016yee}. Further effects from  $\Ks$ \CP violation and matter regeneration have been recently shown to have only a small impact on the \emph{distribution} over the Dalitz plot~\cite{KsCPV}, in contrast to their impact on the global asymmetry. Therefore the \BtoDpi channel can be used to determine the \Fi parameters if the small level of \CP violation in the \Bpm decay is accounted for. 

Pseudoexperiments are performed in which the two \B-decay modes are fit together assuming common \Fi parameters. Independent \xpm and \ypm  observables are required for the two \B decay modes due to different values of the hadronic parameters, \rB and \dB. The value of \rB in \BtoDK is approximately $0.1$, and it is expected that it will be a factor 20 smaller in \BtoDpi decays~\cite{Kenzie:2016yee}. The yields of \BtoDpi are described by a set of equations analogous to Eq.~\eqref{eq:populations}, with the substitutions $\xpmdk \to \xpmdpi$ and $\ypmdk \to \ypmdpi$. An analysis that simultaneously measures the \Fi, \xpmdk, \ypmdk, \xpmdpi, and \ypmdpi parameters is found to be stable only if $\rBDpi>0.03$. At the expected value $\rBDpi=0.005$ the fit is unstable due to high correlations between the \Fi and \xpmdpi and \ypmdpi. Therefore an alternate parameterisation~\cite{Tico:2018qmg,Tico:2019xdx} is introduced, which utilises the fact that \g is a common parameter, and that the \CP violation in \BtoDpi decays can therefore be described by the addition of a single complex variable
\begin{align}
    \xi^{\dpi}=\left(\frac{\rBDpi}{\rBDK}\right)\exp{\left(i\dBDpi-i\dBDK\right)},
\end{align} and in terms of $x_\xi^{\dpi}\equiv\Re (\xi^{\dpi})$ and $y_\xi^{\dpi}\equiv\Im (\xi^{\dpi})$, the $\left(\xpmdpi,\ypmdpi\right)$ parameters are given by
\begin{align}\label{eq:xy_from_xi}
    x_\pm^{\D\pi} &= x_\xi^{\D\pi}\xpmdk - y_\xi^{\D\pi}\ypmdk, 
    & y_\pm^{\D\pi} &= x_\xi^{\D\pi}\ypmdk + y_\xi^{\D\pi}\xpmdk.
\end{align} 
With this parameterisation, the simultaneous fit to \xpmdk, \ypmdk, \xxidpi, \yxidpi (the \CP observables) and \Fi parameters is stable for all values of $\rBDpi$. The simultaneous fit of \BtoDpi and \BtoDK candidates has two advantages. Firstly, the extraction of \Fi in this manner is expected to have negligible associated systematic uncertainty, and reduces significantly the reliance on simulation. Secondly, the \CP-violating observables in \BtoDpi using other \D-decay modes~\cite{LHCb-PAPER-2016-003,LHCb-PAPER-2015-014} are not routinely included in the \g combination of all results because they allow for two solutions of $\left(\rBDpi,\dBDpi\right)$, which makes the statistical interpretation of the full \BtoDh combination problematic~\cite{LHCb-PAPER-2016-032}. The measurement in the \BtoDpi, \DtoKshh decays has the potential to resolve this redundancy, and allow for a more straightforward inclusion of all \BtoDpi results in the combination. A small disadvantage is that the measurement of \g will incorporate information from both \BtoDK and \BtoDpi decay modes and the contribution of each cannot be disentangled. However, since the size of contribution from the \BtoDpi decay to the precision is expected to be negligible in comparison to that from the \BtoDK decay, this is considered an acceptable compromise. 

\begin{figure}[t]
\begin{center}
\includegraphics[width=0.48\textwidth]{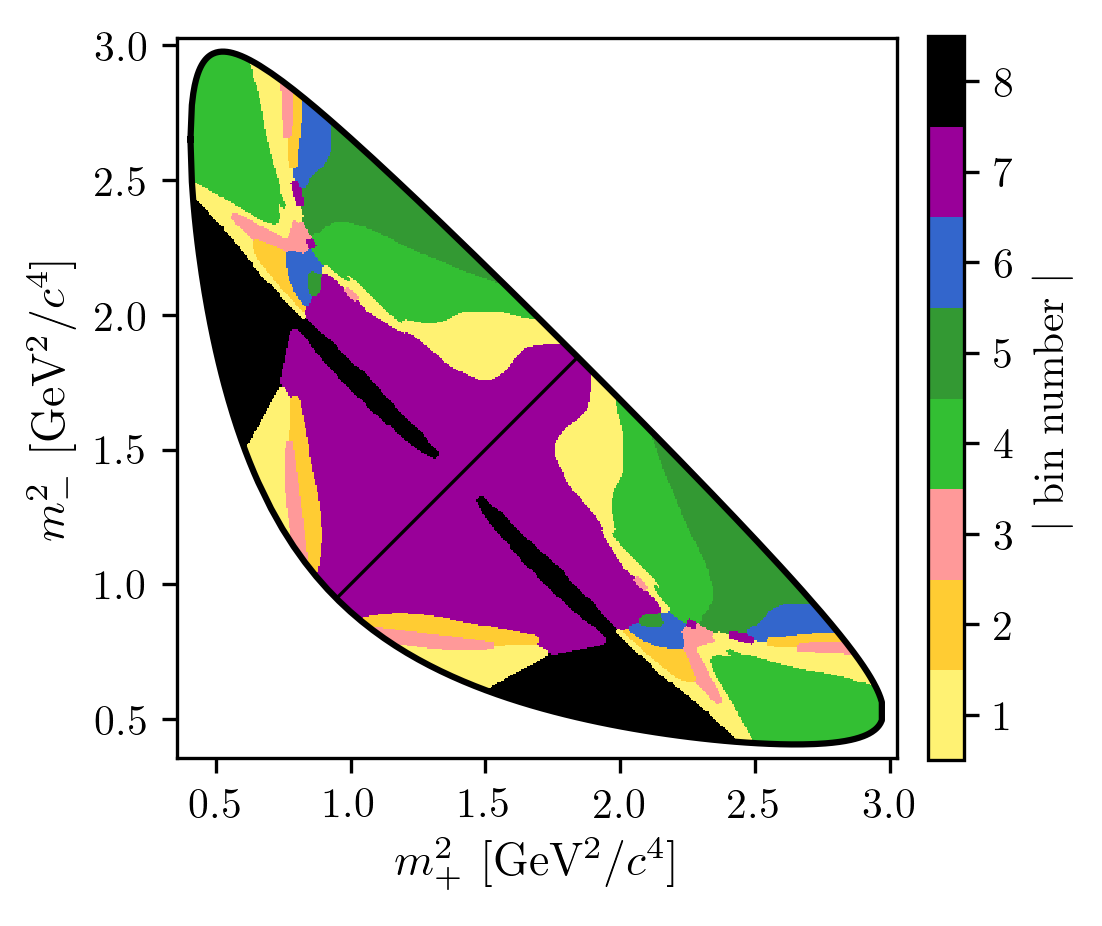}
\includegraphics[width=0.48\textwidth]{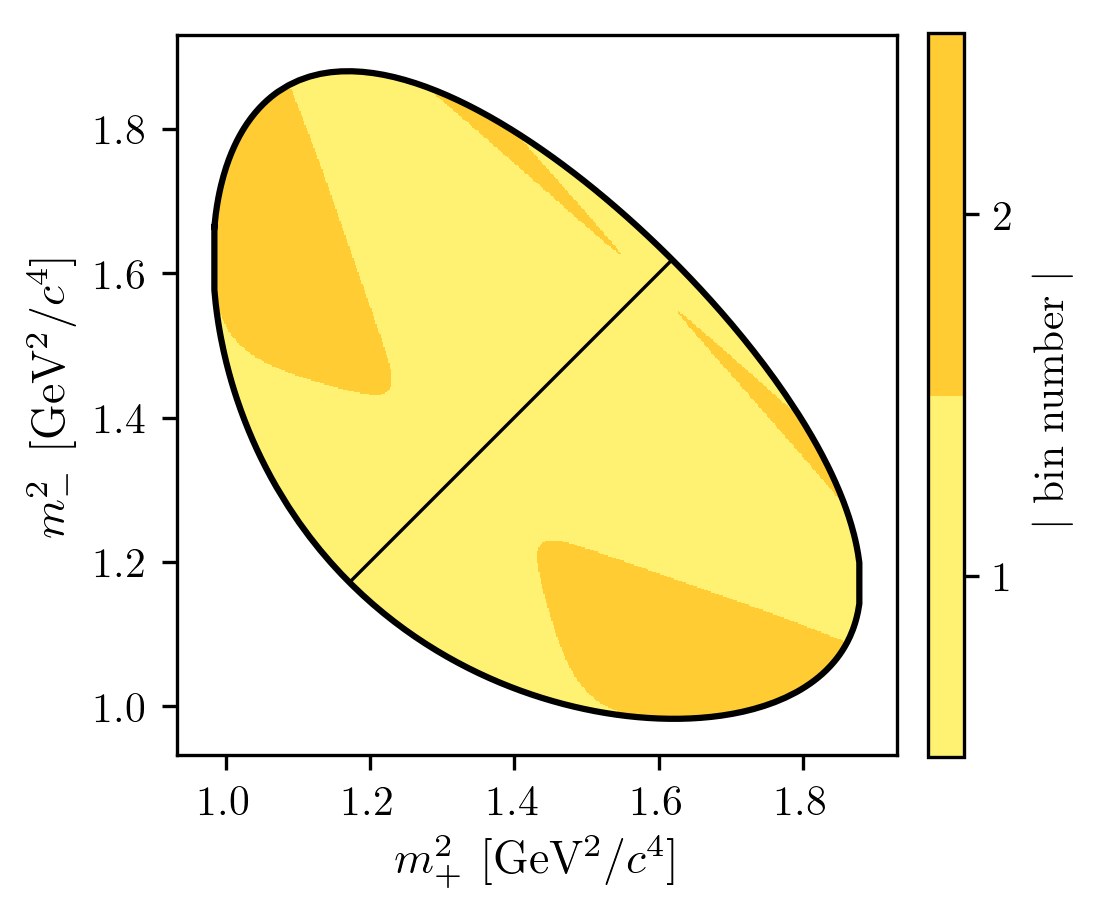}
\caption{\small Binning schemes for (left) \DtoKspp decays and (right) \DtoKskk decays. The diagonal line separates the positive and negative bins, where the positive bins are in the region in which $\msqmin > \msqplus$ is satisfied. }
\label{fig:bins}
\end{center}
\end{figure}

The measurements of $c_i$ and $s_i$ are available in four different $2\times 8$ binning schemes for the \DtoKspp decay.
This analysis uses the scheme called the optimal binning, where the bins have been chosen to optimise the statistical sensitivity to $\gamma$, as described in Ref.~\cite{CLEOCISI}. The optimisation was performed assuming a strong-phase difference distribution as predicted by the \babar model presented in Ref.~\cite{BABAR2008}. For the \KsKK final state, three choices of binning schemes are available, containing $2\times2$, $2\times3$, and $2\times4$ bins.  The guiding model used to determine the bin boundaries
is taken from the \babar study described in Ref.~\cite{BABAR2010}. The \DtoKskk decay mode is dominated by the intermediate $\KS\phi$ and $\KS a(980)$ states which are \CP-odd and \CP-even, respectively, and the narrow $\KS\phi$ resonance is encapsulated within the second bin of the $2\times 2$ scheme. Therefore, most of the sensitivity is encompassed by this scheme, and the additional small gains from the more detailed schemes are offset by low yields and fit instabilities that arise when these bins are used. Therefore, the $2\times2$ bin is used for the analysis of  the \DtoKskk decay mode.  The measurements of $c_i$ and $s_i$ are not biased by the use of a specific amplitude model in defining the bin boundaries. The choice of the model only affects this analysis to the extent that a poor model description of the underlying
decay would result in a reduced statistical sensitivity of the $\gamma$ measurement. The binning choices for
the two decay modes are shown in Fig.~\ref{fig:bins}.

Measurements of the \ci and \si parameters in the optimal binning scheme for the \DtoKspp decay and in the $2\times2$ binning scheme for the \DtoKskk decay are available from both the \cleo and \besiii collaborations. A combination of results from both collaborations is presented in Ref.~\cite{bes3prd} and Ref.~\cite{Krishna} for the \DtoKspp and \DtoKskk decays, respectively. The combinations are used within this analysis.

\section{LHCb Detector}
\label{sec:detector}

The \lhcb detector~\cite{LHCb-DP-2008-001,LHCb-DP-2014-002} is a single-arm forward
spectrometer covering the \mbox{pseudorapidity} range $2<\eta <5$,
designed for the study of particles containing \bquark or \cquark
quarks. The detector includes a high-precision tracking system
consisting of a silicon-strip vertex detector surrounding the $pp$
interaction region, a large-area silicon-strip detector located
upstream of a dipole magnet with a bending power of about
$4{\mathrm{\,Tm}}$, and three stations of silicon-strip detectors and straw
drift tubes placed downstream of the magnet.
The tracking system provides a measurement of the momentum, \ptot, of charged particles with
a relative uncertainty that varies from 0.5\% at low momentum to 1.0\% at 200\gevc. The minimum distance of a track to a primary vertex (PV), the impact parameter (IP), 
is measured with a resolution of $(15+29/\pt)\mum$,
where \pt is the component of the momentum transverse to the beam, in\,\gevc.
Different types of charged hadrons are distinguished using information
from two ring-imaging Cherenkov detectors. Photons, electrons and hadrons are identified by a calorimeter system consisting of
scintillating-pad and preshower detectors, an electromagnetic
and a hadronic calorimeter. Muons are identified by a
system composed of alternating layers of iron and multiwire
proportional chambers.
The online event selection is performed by a trigger, 
which consists of a hardware stage, based on information from the calorimeter and muon
systems, followed by a software stage, which applies a full event
reconstruction. The events that are selected for the analysis either have  final-state tracks of the signal decay that are subsequently associated with an energy deposit in the calorimeter system that satisfies the hardware stage trigger, or are selected because  one of the other particles in the event, not reconstructed as part of the signal candidate, fulfils any hardware stage trigger requirement.
At the software stage, it is required that at least one particle should have high \pt and high \chisqip, where \chisqip is defined as the difference in the primary vertex fit \chisq with and without the inclusion of that particle. A multivariate algorithm~\cite{Gligorov:2012qt} is used to identify secondary vertices consistent with being a two-, three-, or four-track $b$-hadron decay. The PVs are fitted with and without the tracks of the decay products of the $B$ candidate, and the PV that gives the smallest \chisqip is associated with the $B$ candidate.

Simulation is required to model the invariant-mass distributions of the signal and background contributions and determine the selection efficiencies of the background relative to the signal decay modes. It is also used to provide an approximation for the efficiency variations over the phase space of the \D decay for systematic studies. In the simulation, $pp$ collisions are generated using
  \pythia~\cite{Sjostrand:2007gs,*Sjostrand:2006za} with a specific \lhcb configuration~\cite{LHCb-PROC-2010-056}.
  Decays of unstable particles
  are described by \evtgen~\cite{Lange:2001uf}, in which final-state
  radiation is generated using \photos~\cite{Golonka:2005pn}. The decays \DtoKspp and \DtoKskk are generated uniformly over phase space.
  The interaction of the generated particles with the detector, and its response,
  are implemented using the \geant
  toolkit~\cite{Allison:2006ve, *Agostinelli:2002hh} as described in
  Ref.~\cite{LHCb-PROC-2011-006}. With the exception of the signal decay, the
  simulated event is reused
  multiple times~\cite{LHCb-DP-2018-004}. Some subdominant backgrounds are generated with a fast simulation~\cite{Cowan:2016tnm} that can mimic the geometric acceptance and tracking efficiency of the \lhcb detector as well as the dynamics of the decay.

\section{Selection}
\label{sec:selection}

The selection closely follows that of Ref.~\cite{LHCb-PAPER-2018-017}.
Decays of \decay{\KS}{\pip\pim} are reconstructed in two different ways:
the first involving \KS mesons that decay early enough for the
pions to be reconstructed in the vertex detector; and the
second containing \KS that decay later such that track segments of the
pions cannot be formed in the vertex detector. The first and second types of reconstructed \KS decays are
referred to as \emph{long} and \emph{downstream} candidates, respectively. The
\emph{long} candidates have the best mass, momentum and vertex resolution, but approximately two-thirds of the signal candidates belong to the \emph{downstream} category. 

The \D meson candidates are built by combining a \KS candidate with two tracks assigned either the pion or kaon hypothesis. A \B candidate is then formed by combining the \D meson candidate with a further track. At each stage of combination, selection requirements are placed to ensure good quality vertices, and \KS and \D candidate invariant-masses are required to be close to their nominal mass~\cite{PDG2020}. Mutually exclusive particle identification (PID) requirements are placed on the companion track from the \B decay to separate \BtoDK and \BtoDpi candidates, where the companion refers to the final state \pipm or \Kpm meson produced in the \BtoDh decay. PID requirements are also placed on the charged decay products of the \D meson to reduce combinatorial background. A series of selection requirements are placed on the candidates to remove background from other \B meson decays. 
A background from \BtoDh decays where the \D meson decays to either $\pip\pim\pip\pim$ or $\Kp\Km\pip\pim$ is rejected by requiring that the \emph{long} \KS candidates decay a significant distance from the \D vertex. Similarly, the \D meson is required to have travelled a significant distance from the \B vertex to suppress \B decays with the same final state, but where there is no intermediate \D meson decay. 
Semileptonic decays of the type $\Dz \to K^{*-} l^+\nu$, where charge-conjugate decays are implied, can be reconstructed as \DtoKshh with expected contamination rates of the order of a percent. To suppress electron to pion misidentification, a veto is placed on the pion from the \D decay that has the opposite charge with respect to the companion particle, if the PID response suggests it is an electron. To suppress the similar muonic background, it is required that the charged track from the \D decay has no corresponding activity in the muon detector. This veto also suppresses signal decays where the pion or kaon meson decays before reaching the muon detector. Therefore, it is  applied on both charged tracks from the \D decay, as these events have a worse resolution on the Dalitz plot, which is undesirable. Finally, the same requirement is placed on the companion track to suppress $\B \to \D \mu \nu$ decays.

The large remaining combinatorial background is suppressed through the use of a boosted decision tree (BDT)~\cite{Breiman,AdaBoost}  multivariate classifier. The BDT is trained on simulated signal events. The background training sample is obtained from the far upper sideband of the $m(Dh^\pm)$ mass distribution between 5800-7000\mevcc, in order to provide a sample independent from the data which will be used in the fit to determine the \CP observables. A separate BDT is trained for \B decays containing \emph{long} or \emph{downstream} \KS candidates. The input variables given to each BDT include momenta of the \B, \D, and companion particles, the absolute and relative positions of decay vertices, as well as parameters that quantify the fit quality in the reconstruction; the parameter set is identical to the one used in the previous \lhcb measurement and listed in detail in Ref.~\cite{LHCb-PAPER-2018-017}. The BDT has been proven not to bias the $m(Dh^\pm)$ distribution. A series of pseudoexperiments are run to find the threshold values for the two BDTs which provide the best sensitivity to \Pgamma.  This requirement rejects approximately 98$\%$ of the combinatorial background that survives all other selection requirements, while having an efficiency of approximately 93$\%$ in simulated \BtoDK decays. The selection applied to \BtoDK and \BtoDpi candidates is identical between the two decay modes with the exception of the PID requirement on the companion track.

\begin{figure}
    \centering
    \includegraphics[width=0.45\columnwidth]{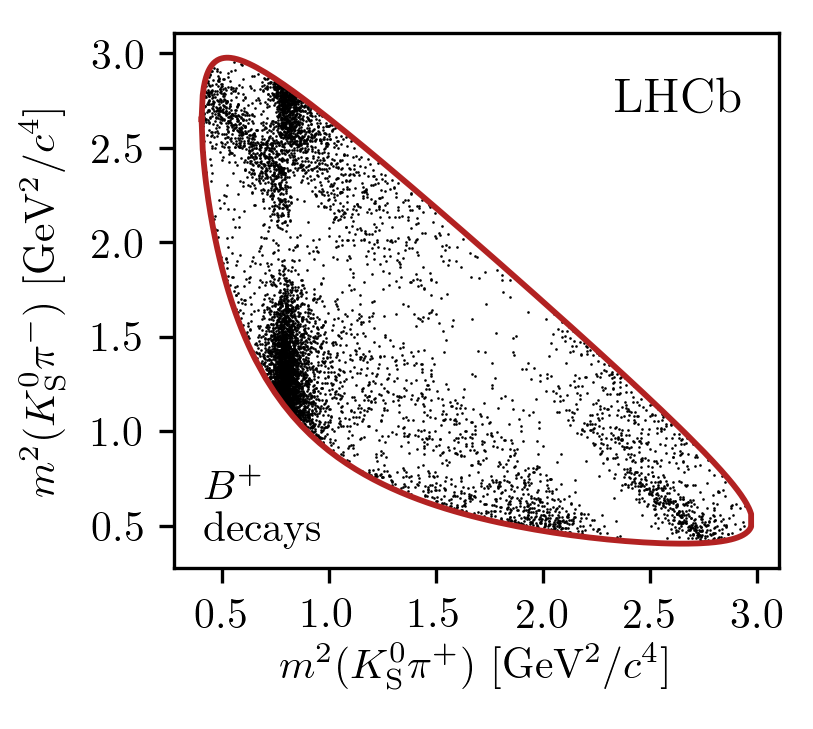}
    \includegraphics[width=0.45\columnwidth]{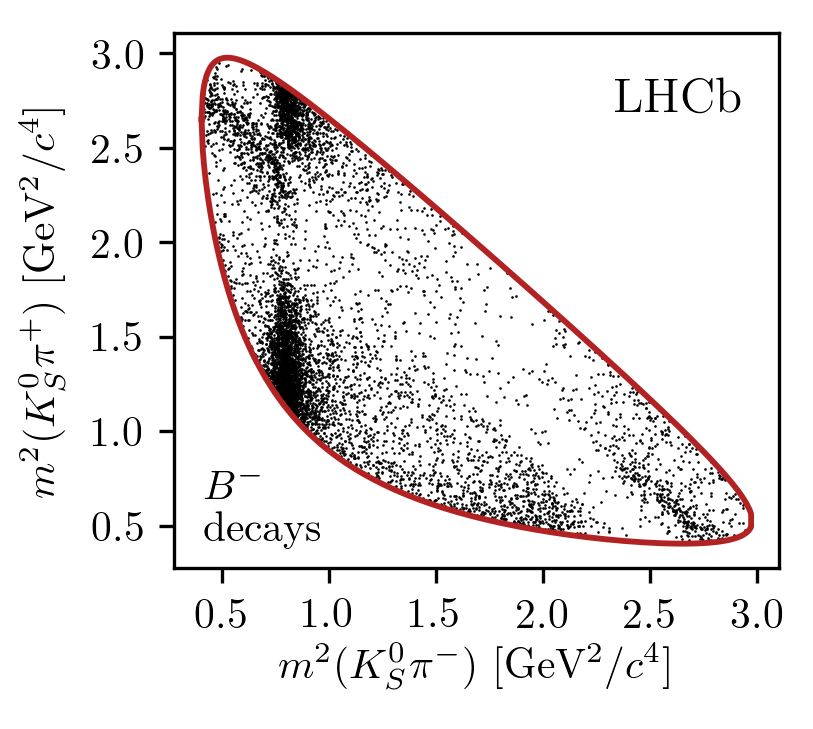}
 \includegraphics[width=0.45\columnwidth]{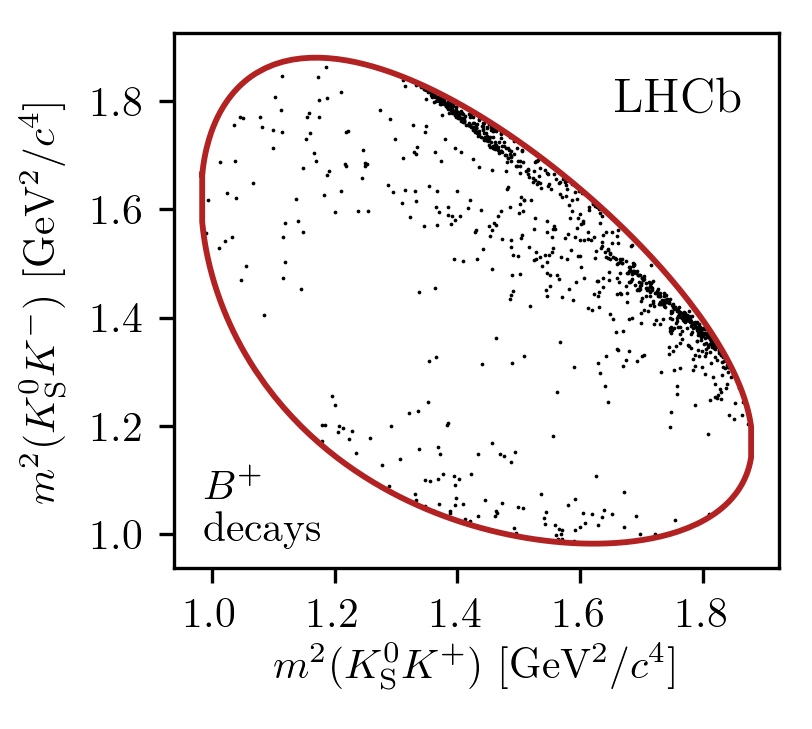}
    \includegraphics[width=0.45\columnwidth]{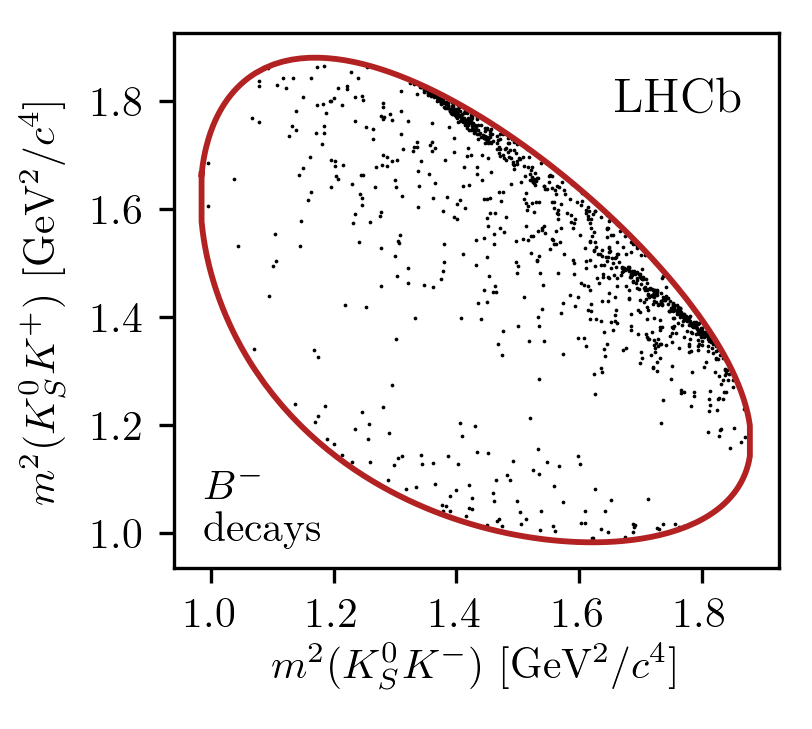}
    \caption{Dalitz plot for $\D$ decays of (left) $\Bp\to\D\Kp$ and (right) $\Bm\to\D\Km$ candidates in the signal region, in the (top) \DtoKspipi and (bottom) \DtoKsKK channels. The horizontal and vertical axes are interchanged between the \Bp and \Bm decay plots to aid visualisation of the \CP asymmetries between the two distributions.}
    \label{fig:dalitz_plots_DK}
\end{figure}

A signal region is defined as within 30\mevcc of the \B-meson mass~\cite{PDG2020}. The phase-space distributions for candidates in this range are shown in the Dalitz plots of Fig.~\ref{fig:dalitz_plots_DK} for \BtoDK candidates. The data are split by the final state of the \D decay and by the charge of the \B meson. Small differences between the phase-space distributions in $\Bp\to\D\Kp$ and $\Bm\to\D\Km$ decays are visible in the \KsPiPi final state. 

\section{The \boldmath{$D K$} and \boldmath{$\D \pi$} invariant-mass spectra}
\label{sec:massfit}

The analysis uses a two-stage  strategy to determine the \CP observables. First, an extended maximum-likelihood fit to the invariant-mass spectrum of all selected \Bpm candidates in the mass range 5080 to 5800 \mevcc is performed, with no partition of the \D phase space. This fit is referred to as the \emph{global} fit. The global fit is used to determine the signal and background component parameterisations, which are subsequently used in a second stage where the data are split by \B charge and partitioned into the Dalitz plot bins to determine the \CP observables.

\begin{figure}
    \centering
    \includegraphics[width=\columnwidth]{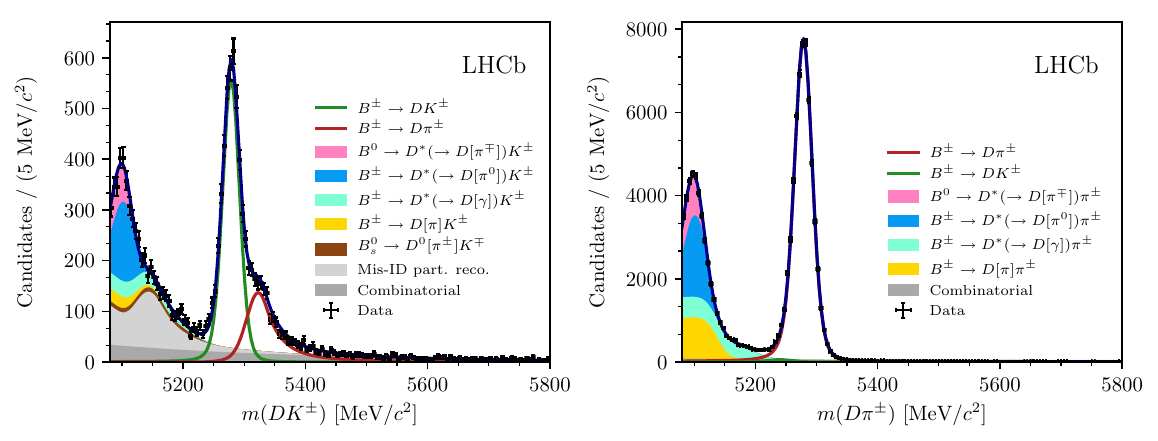}
    \includegraphics[width=\columnwidth]{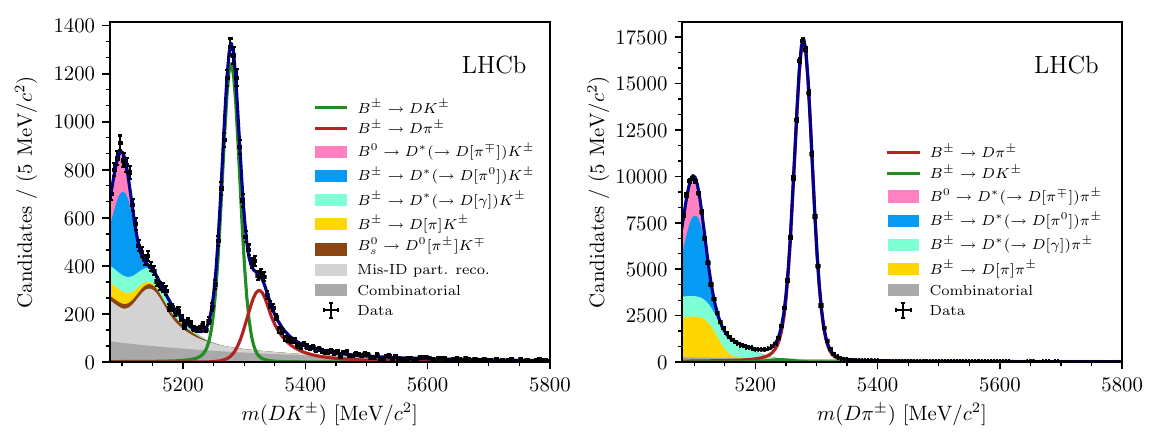}
    \caption{Invariant mass distributions for the (left) \BtoDK channel and (right) \BtoDpi channel with \DtoKspp. The top (bottom) plots show data where the \KS candidate is \emph{long} (\emph{downstream}). Square brackets in the legend denote a particle that has not been reconstructed. }
    \label{fig:kspipimass}
\end{figure}

\begin{figure}
    \centering
    \includegraphics[width=\columnwidth]{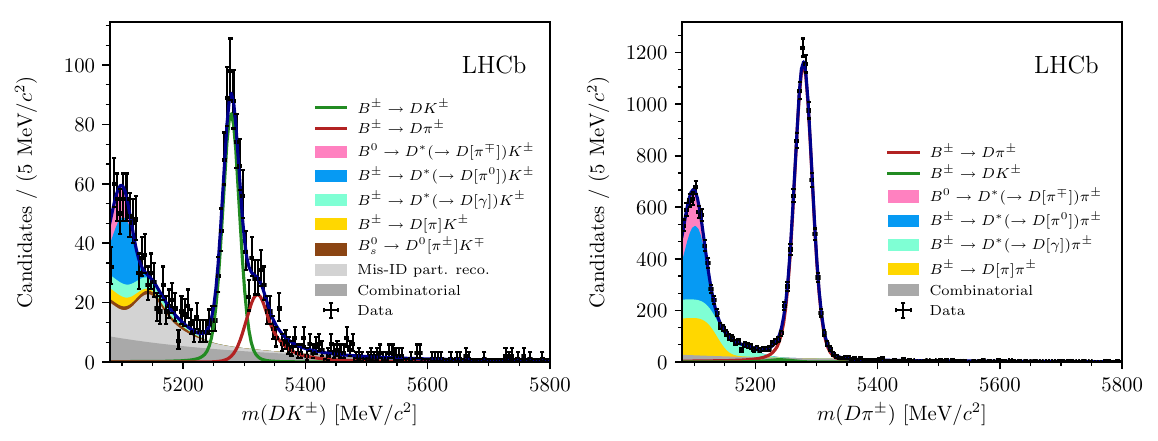}
    \includegraphics[width=\columnwidth]{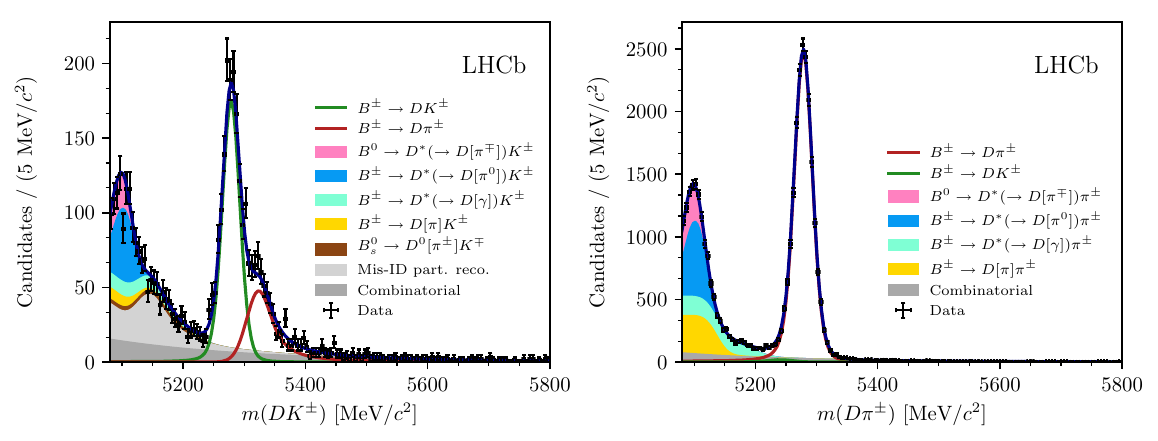}
    \caption{Invariant mass distributions for the (left) \BtoDK channel and (right) \BtoDpi channel with \DtoKskk. The top (bottom) plots show data where the \KS candidate is \emph{long} (\emph{downstream}). Square brackets in the legend denote a particle that has not been reconstructed.}
    \label{fig:kskkmass}
\end{figure}

The invariant mass distributions of the selected \Bpm candidates are shown for \DtoKspp and \DtoKskk candidates in Figs.~\ref{fig:kspipimass} and~\ref{fig:kskkmass}, respectively, together with the results of the global fit superimposed. The invariant mass is kinematically constrained through a fit imposed on the full \Bpm decay chain~\cite{Hulsbergen:2005pu}. The \D and \KS candidates are constrained to their known masses\cite{PDG2020} and the \Bpm candidate momentum vector is required to point towards the associated PV. The data sample is split into 8 categories depending on the reconstructed \B decay, \D decay mode, and \KS category, since the latter exhibits slightly different mass resolutions. The fit is performed simultaneously for all categories in order to allow parameters to be shared. 

The peaks centered around 5280\mevcc correspond to the signal \BtoDK and \BtoDpi candidates.  The  parameterisation for the signal invariant-mass shape is determined from simulation; the invariant-mass distribution is modelled with a sum of the probability density function (PDF) for a Gaussian distribution, $f_G(m|m_B, \sigma)$, and  a modified Gaussian PDF that is used to account for the radiative tail and the wider resolution of signal events that are poorly reconstructed. The modified Gaussian has the form
\begin{align}
    f_\mathrm{MG}(m|m_B, \sigma, \alpha_L, \alpha_R, \beta)\propto \left\{
    \begin{array}{ll}
    \exp\left[\frac{-\Delta m^2 (1+\beta \Delta m^2)}{2\sigma^2 + \alpha_L \Delta m^2}\right], & \Delta m = m - m_B  < 0 \\ [8pt]
    \exp\left[\frac{-\Delta m^2 (1+\beta \Delta m^2)}{2\sigma^2 + \alpha_R \Delta m^2}\right], & \Delta m = m - m_B  > 0,
    \end{array} 
    \right.,
\end{align}
which is Gaussian when $\Delta m^2 \ll \sigma^2/\alpha_{L/R}$ or $\Delta m^2 \gg \beta^{-1}$ (with widths of $\sigma$ and $\sqrt{\alpha_{L/R}/\beta}$, respectively), with an exponential-like transition that is able to model the effect of the experimental resolution of \lhcb. Thus, the signal PDF has the form
\begin{align}
    f_\mathrm{signal}(m|m_B, \sigma, \alpha_L, \alpha_R, \beta, k) =
    k\cdot
    &f_\mathrm{MG}(m|m_B, \sigma, \alpha_L, \alpha_R, \beta) \notag \\
    &+ (1-k) \cdot f_\mathrm{G}(m|m_B, \sigma)
\end{align}
The values of the tail parameters ($\alpha_L$, $\alpha_R$, \Pbeta) and $k$ are fixed from simulation and are common for the two \D decays (which is possible due to the applied kinematic constraints) but different for each \B decay and type of \KS candidate. The signal mass, $m_B$, is determined in data and is the same for all categories. The width, $\sigma$, of the signal PDF is determined by the data and allowed to be different for each $B$ decay and type of \KS candidate. The width is narrower in \BtoDK decays compared to \BtoDpi decays due to the smaller free energy in the decay. The width is approximately 3\% narrower in decays with \emph{long} \KS candidates. The signal yield is determined in each of the categories where the candidates are reconstructed as \BtoDpi. The signal yield in the corresponding category where the candidates are reconstructed as \BtoDK is determined by multiplying the \BtoDpi yield by the parameters $\mathcal{B}\times\epsilon$. The parameter $\mathcal{B}$ corresponds to the ratio of the branching fractions for \BtoDK and \BtoDpi decays, while the correction factor, $\epsilon$, takes into account the ratio of PID and selection efficiencies, and is determined for each pair of \BtoDK and \BtoDpi categories and fixed in the fit. The parameter $\mathcal{B}$ is shared across all categories and is found to be consistent with Ref.~\cite{PDG2020}.

To the right of the \BtoDK peak there is a visible contribution from \BtoDpi decays that are reconstructed as \BtoDK decays. The corresponding contribution in the \BtoDpi category is minimal due to the smaller branching fraction of \BtoDK, but is accounted for in the fit. The rates of these cross-feed backgrounds are fixed from PID efficiencies determined in calibration data, which is reweighted to match the momentum and pseudorapidity distributions of the  companion track of the signal. A data-driven approach is used to determine the PDF of \BtoDpi decays that are reconstructed as \BtoDK candidates, by using \BtoDpi decays and recalculating the invariant mass when the Kaon hypothesis is applied to the companion particle. Full details of the procedure are described in Ref.~\cite{LHCb-PAPER-2018-017}. The same procedure is implemented to determine the PDF of \BtoDK decays reconstructed as \BtoDpi candidates. 

The background observed at invariant masses smaller than the signal peak are candidates that originate from other \B-meson decays where not all decay products have been reconstructed. Due to the selected invariant-mass range it is only necessary to consider \B meson decays where a single photon or pion has not been reconstructed. This background type is split into three sources; the first where the candidate originates from a \Bpm or \Bz meson, referred to as partially reconstructed background, the second where the candidate originates from a \Bs meson, and the third where the candidate originates from a \Bpm or \Bz and furthermore one of the reconstructed tracks is assigned the kaon hypothesis, when the true particle is a pion. The latter type of background appears in the \BtoDK candidates and is referred to as misidentified partially reconstructed background. The corresponding type of background is not modelled in the \BtoDpi candidates, since it is suppressed due to the branching fractions involved and the majority is removed by the lower invariant-mass requirement. 

There are contributions from $\Bz \to D^{*\pm} h^\mp$ and $\Bpm \to \D^{*0} h^\pm$ decays in all categories, where the pion or the photon originating from the $\D^{*}$ meson is not reconstructed. The invariant-mass distributions of these decays depend on the spin and mass of the missing particle as described in Ref.~\cite{LHCb-PAPER-2017-021}. The parameters of these shapes are determined from simulation, with the exception of a free parameter in the fit to characterise the resolution. The decays $\B^{\pm, 0} \to \D \pipm \pi^{0,\mp}$ contribute to the \BtoDpi candidates where one of the pions from the \B decay is not reconstructed. The shape of this background is determined from simulated $\Bpm \to \D \rho^\pm$ and $\Bz \to D\rho^0$ decays. 
The decays $\Bpm \to \D \Kpm \piz$ and $\Bz \to \D \Kp \pim$ contribute to the \BtoDK candidates where the pion is not reconstructed. The invariant-mass distribution for these events is based on the amplitude model of $\Bz \to \D \Kp\pim$ decays~\cite{LHCb-PAPER-2015-059}. The model is used to generate four-vectors of the decay products, which are smeared to account for the LHCb detector resolution. The invariant mass is then calculated omitting the particle that is not reconstructed, and this distribution is subsequently fit to determine the fixed distribution for the fit. The same shape is used for the $\Bpm \to \D \Kpm \piz$ decay as the corresponding amplitude model is not available. 
Finally, the \BtoDK candidates also have a contribution from $\Bs \to \Dzb \pip\Km$ decays where the pion is not reconstructed. The shape of this contribution is determined in a similar manner to that of $\Bz \to \D \Kp\pim$ decays using the $\Bs \to \Dzb \pip\Km$ amplitude model determined in Ref.\cite{LHCb-PAPER-2014-036}.

The yield of the partially reconstructed background is a free parameter in each \BtoDpi sample and related to the yield in the corresponding \BtoDK sample via the free parameter $\mathcal{B_L}$ and correction factors from PID and selection efficiencies. Analogously to the signal-yield parameterisation, $\mathcal{B_L}$ is a single parameter, common to all categories, but in this case has no direct physical meaning. The relative yield of $\Bpm \to D^{*}(\to \D [\gamma])\pipm$ and $\Bz \to D^{*}(\to \D[\pimp])\pipm$ decays, where the particle within the square brackets is the one not reconstructed, are fixed from branching fractions~\cite{PDG2020}, and selection efficiencies determined from simulation. The fractional yields of $\Bpm \to D^{*0}(\to \D [\gamma])\pipm$, and $\B^{\pm,0} \to D [\pi^{0,\mp}]\pipm$ decays are determined in the fit and are constrained to be the same for each \BtoDpi sample. Due to the lower yields in the \BtoDK category and presence of additional backgrounds, the relative fractions of the various \Bpm and \Bz components are all fixed using information from branching fractions~\cite{PDG2020} and selection efficiencies from simulation. The yield of the $\Bs \to \Dzb \pip \Km$ decays is fixed relative to the yield of \BtoDpi decays in the corresponding category using branching fractions~\cite{PDG2020}, the fragmentation fraction~\cite{LHCB-PAPER-2018-050}, and relative selection efficiencies. Hence in total there are six free parameters to determine the partially reconstructed background yields. Four of these measure the yield in each \BtoDpi sample and a further single parameter determines the yield in the corresponding \BtoDK sample. A final floating parameter determines the relative yield between two of the background components. 

The shapes for the misidentified partially reconstructed backgrounds are determined from simulation, weighted by the PID efficiencies from calibration data. The yield of these backgrounds are determined from the partially reconstructed yields in the \BtoDpi candidates, and the relative selection efficiencies, which include the PID efficiencies from calibration data and the selection efficiency due to requiring the reconstructed invariant mass to be above 5080\mevcc.  
The final component of background is combinatorial which is parameterised by an exponential function. The yield and slope of this background in each category are free parameters. The yields of the different signals and background types are integrated in the signal region 5249--5309\mevcc and reported in Table~\ref{tab:fit_yields}. The \BtoDK yields in categories of different \D decay and type of \KS candidate have uncertainties that are smaller than their Poisson uncertainty since they are determined using the value of $\mathcal{B}$, which is measured from all \BtoDK candidates. 

\begin{table}[]
    \centering
        \caption{The signal and background yields in the region $m_B\in[5249, 5309]\mevcc$ as obtained in the fit. For the \BtoDK candidates, the yield of the partially reconstructed background includes the contributions from \Bs decays and misidentified partially reconstructed backgrounds.}
    \label{tab:fit_yields}
    \footnotesize

\begin{tabular}{ll|cc|cc}
\toprule
\ & Reconstructed as: & \multicolumn{2}{c|}{$\BtoDK$} & \multicolumn{2}{c}{$\BtoDpi$} \\
\midrule
\D decay & Component & \emph{long} & \emph{downstream} &\emph{long} & \emph{downstream} \\
\midrule
$\DtoKspipi$ & $\Bpm\to\D\Kpm$             & $ 3798 \pm    41$& $ 8735 \pm    89$& $  182 \pm     3$& $  433 \pm     8$ \\
             & $\Bpm\to\D\pipm$            & $  342 \pm     3$& $  691 \pm     5$& $55096 \pm   240$& $124786 \pm   368\phantom{1}$ \\
             & Part. reco. background      & $  114 \pm     3$& $  246 \pm     6$& $   \phantom{1}36 \pm     7$& $   \phantom{11}81 \pm    12$ \\
             & Combinatorial               & $  \phantom{1}206 \pm    36$& $  \phantom{1}458 \pm    60$& $  \phantom{1}392 \pm    66$& $ \phantom{1}1142 \pm   127$ \\
\midrule
$\DtoKskk$   & $\Bpm\to\D\Kpm$             & $  576 \pm     8$& $ 1203 \pm    15$& $   \phantom{1}29 \pm     1$& $  \phantom{1} 61 \pm     1$ \\
             & $\Bpm\to\D\pipm$            & $   \phantom{1}56 \pm     1$& $  104 \pm     2$& $ 8196 \pm    92$& $17863 \pm   137$ \\
             & Part. reco background      & $   \phantom{1}17 \pm     2$& $   \phantom{1}34 \pm     2$& $   \phantom{11} 5 \pm     3$& $  \phantom{1} 11 \pm     5$ \\
             & Combinatorial               & $   \phantom{11}44 \pm    13$& $  \phantom{11} 75 \pm    20$& $  \phantom{1}127 \pm    32$& $ \phantom{1} 288 \pm    52$ \\
\bottomrule
\end{tabular}

\end{table}

\section{\boldmath{$\CP$} observables}
\label{sec:cpfit}

To determine the \CP observables the data are divided into 16 categories (\B decay, \B charge, \D decay, type of \KS candidate) and then further split into each Dalitz plot bin. A simultaneous fit to the invariant-mass distribution is performed in all categories and Dalitz plot bins. The mass shape parameters are all fixed from the global mass fit. The lower limit of the invariant mass is increased to 5150\mevcc to remove a large fraction of the partially reconstructed background.  The signal yield in each bin is parameterised using Eq.~\eqref{eq:populations} or the analogous set of expressions for \BtoDpi. These equations are normalised such that the parameters $h_{\Bpm}$ represent the total observed signal yield in each category, and these are measured independently. The yield of misidentified \BtoDpi (\BtoDK) decays in each \BtoDK (\BtoDpi) sample is determined by the PID efficiencies and the signal yield in the corresponding \BtoDpi (\BtoDK) sample.

The parameters \xpmdk, \ypmdk, \xxidpi, and \yxidpi are free parameters in the fit and common to the \KS and \D decay categories. The parameters \ci and \si are fixed to those determined from the combination of \besiii and \cleo data~ in Ref.\cite{bes3prd} for the \DtoKspp decays and in Ref.~\cite{Krishna} for the \DtoKskk decays. The $F_i$ parameters for each \D decay are determined in the fit; separate sets of \Fi parameters are determined for the two types of \KS candidates because the efficiency profile over the Dalitz plot differs between the \KS selections. Since the $F_i$ parameters must satisfy the constraints $\sum_i F_i = 1,\,\Fi\in[0,1]$, the fit can suffer from instability if they are included in a naive way due to large correlations. Therefore, the $F_i$ parameters are reparameterised as a series of recursive fractions with parameters, $\mathcal{R}_i$, determined in the fit. The relation between the $F_i$ and $\mathcal{R}_i$ parameters is given by
\begin{align}
     F_i = \left\{
     \begin{array}{ll}
         \mathcal R_i  &,\quad i = -\mathcal N \\
         \mathcal R_i \prod_{j < i}(1-\mathcal R_j) &,\quad -\mathcal N < i < +\mathcal N \\
         \phantom{R_i}\prod_{j < i}(1-\mathcal R_j) &,\quad i = +\mathcal N.
     \end{array}
   \right.,
 \end{align}
for a binning scheme with $2\times\mathcal N$ bins.

The yield of the combinatorial background in each bin is a free parameter. The yield of the partially reconstructed background from \Bpm or \Bz decays in the \BtoDpi and \BtoDK samples is also a free parameter in each bin. The composition of the partially reconstructed background is determined from the global fit described in Sect.~\ref{sec:massfit}, taking into account the change in the lower limit of invariant mass. The yield of the misidentified partially reconstructed background in the \BtoDK samples is determined via the background yield in the corresponding \BtoDpi bin and the relative PID and selection efficiencies. The yield of the $\Bs \to \Dzb \Km \pip$ background is fixed from the global fit and is divided into the Dalitz plot bins according to the $F_i$ such that it has the distribution of a \Dz decay in the $B^+$ categories and the distribution of a \Dzb decay in the $B^-$ categories. 

There is a small fraction of bins where either the partially reconstructed background or combinatoric background yield is less than one. These bins are identified in a preliminary fit and the background yield is fixed to zero. This procedure is carried out to improve the fit stability. 

Pseudoexperiments are performed to investigate any potential biases or remaining instabilities in the fit. The candidate yields and mass distributions in these pseudoexperiments are based on the global fit results. The pull distributions are well described by a Gaussian function and are found to have mean and width consistent with 0 and 1, respectively. 

\begin{figure}
    \centering
    \includegraphics{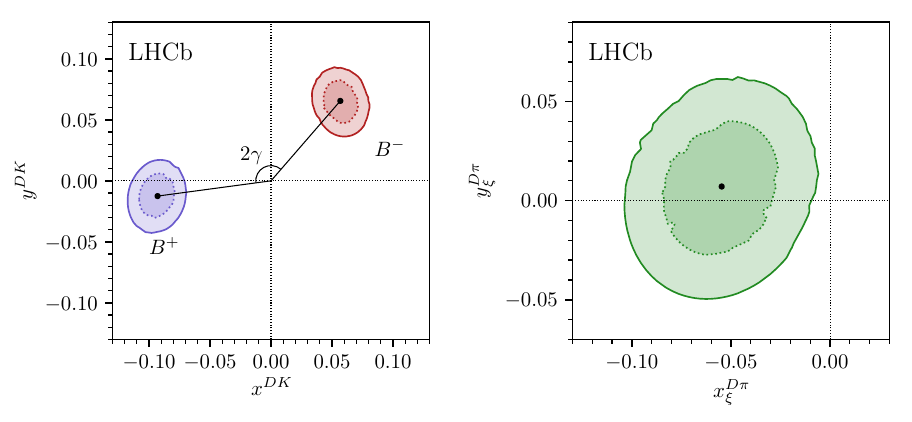}
    \caption{Confidence levels at 68.2\,\% and 95.5\,\% probability for (left, blue) $(\xp^{\D\Kpm},\yp^{\D\Kpm})$, (left, red) $(\xm^{\D\Kpm}, \ym^{\D\Kpm})$, and (right, green) $(x_\xi^{\D\pipm},y_\xi^{\D\pipm}) $ as measured in \BtoDK and \BtoDpi decays with a profile likelihood scan. The black dots show the central values }
    \label{fig:sunnysideup}
\end{figure}

The results for \xpmdk, \ypmdk, \xxidpi, and \yxidpi are presented in Fig.~\ref{fig:sunnysideup} along with their likelihood contours, where only statistical uncertainties are considered. The two vectors defined by the origin and the end-point coordinates $(\xmdk,\ymdk)$ and $(\xpdk,\ypdk)$ give the values for \rBDK for \Bm and \Bp decays. The signature for \CP violation is that these vectors must have non-zero length and have a non-zero opening angle between them, since this angle is equal to $2\gamma$, as illustrated on the figure. Therefore, the data exhibit unambiguous features of \CP violation as expected. The relation between the hadronic parameters in \BtoDpi and \BtoDK decays is also illustrated in Fig~\ref{fig:sunnysideup}, where the vector defined by the coordinates (\xxidpi,\yxidpi) is the relative magnitude of $r_B$ between the two decay modes. It is consistent with the expectation of 5\% ~\cite{Kenzie:2016yee}. The normalization constants give global asymmetries that are consistent with the expectation of asymmetries from production, detection and neutral kaon effects. 

A series of cross checks is carried out by performing separate fits by splitting the data sample into data-taking periods by year, type of \KS candidate, \D-decay, hardware trigger path, and magnet polarity. The results are consistent between the datasets. As an additional cross check, the two-stage fit procedure is repeated with a number of different selections applied to the data. Of particular interest are the alternative selections that significantly affect the presence of specific backgrounds: the fits where the value of the BDT threshold is varied to decrease the level of combinatorial background and those where the choice of PID selection is changed to result in a substantially lower level of misidentified \BtoDpi decays and misidentified partially reconstructed background in the \BtoDK candidates. The variations in the central values for the \CP observables are consistent within the statistical uncertainty associated with the change in the data sample.

\begin{figure}
    \centering
    \includegraphics[width=0.45\columnwidth]{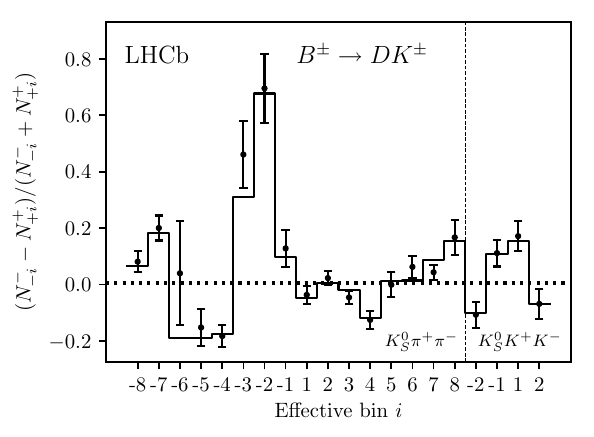}
    \includegraphics[width=0.45\columnwidth]{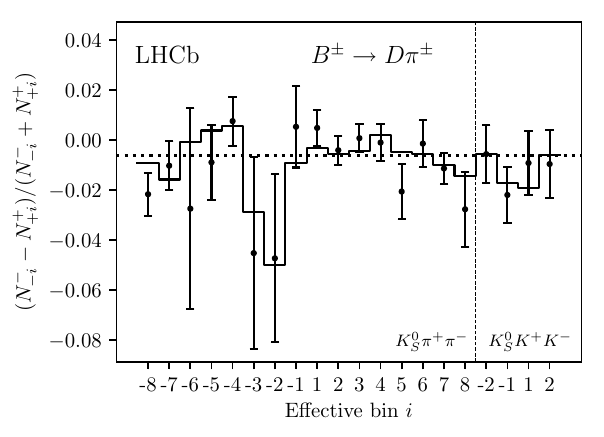}
    \caption{The bin-by-bin asymmetries $(N^-_{-i}-N^+_{+i})/(N^-_{-i}+N^+_{+i})$ for each Dalitz-plot bin number for (left) \BtoDK decays and (right) \BtoDpi decays. The prediction from the central values of the \CP-violation observables is shown with a solid line and the asymmetries obtained in fits with independent bin yields are shown with the error bars. The predicted asymmetries in a fit that does not allow for \CP violation are shown with a dotted line. The vertical dashed line separates the \Kspipi and \KsKK bins on the horizontal axis.}
    \label{fig:asymmetries}
\end{figure}

In order to assess the goodness of fit and to demonstrate that the equations involving the \CP parameters provide a good description of the signal yields in data, an alternative fit is performed where the signal yield in each \BtoDK and \BtoDpi bin is measured independently. The alternate fit is performed simultaneously in all categories in order to correctly determine the yield of misidentified candidates. These yields are compared with those predicted from the values of $(\xpmdk,\ypmdk)$ in the default fit and a high level of agreement is found. In order to visualise the observed \CP violation, the asymmetry, $(N^-_{-i}-N^+_{+i})/(N^-_{-i}+N^+_{+i})$, is computed for \emph{effective bin pairs}, defined to comprise bin $i$ for a \Bp decay and bin $-i$ for a \Bm decay. Figure~\ref{fig:asymmetries} shows the obtained asymmetries and those predicted by the values of the \CP observables obtained in the fit. A further fit that does not allow for \CP violation is carried out by imposing the conditions \xpdk= \xmdk, \ypdk= \ymdk. This determines the predicted asymmetry arising from detector and production effects. In the \BtoDK sample the \CP violation is clearly visible as the data are inconsistent with the \CP-conserved hypothesis. The observed asymmetries correspond to a $\sim 10\sigma$ deviation given the expectation in the \CP-conserving scenario. The predicted asymmetries in the \BtoDpi decay are an order of magnitude smaller. The data in this analysis cannot distinguish between the \CP-violating and \CP-conserving predictions for \BtoDpi due to the relatively large statistical uncertainties. 

\section{Systematic uncertainties}
\label{sec:syst}

\begin{table}
\centering
\caption{Overview of all sources of uncertainty, $\sigma$, on \xpmdk, \ypmdk, \xxidpi, and \yxidpi. All uncertainties are quoted $\times 10^{-2}$.
\label{tab:systematic_uncertainties}}

\footnotesize

\footnotesize

\begin{tabular}{l|cccccc}
\toprule
Source & 
$\sigma(\xmdk)$ & $\sigma(\ymdk)$ & 
$\sigma(\xpdk)$ & $\sigma(\ypdk)$ &
$\sigma(\xxidpi)$ & $\sigma(\yxidpi)$ \\
\midrule
Statistical                              & 0.96  & 1.14  & 0.98  & 1.23  & 1.99  & 2.33  \\ 
\midrule
Strong-phase inputs                      & 0.23  & 0.35  & 0.18  & 0.28  & 0.14  & 0.18  \\ 
\midrule
Efficiency correction of $(c_i, s_i)$    & 0.11  & 0.05  & 0.05  & 0.10  & 0.08  & 0.09  \\ 
Mass-shape parameters                    & 0.05  & 0.08  & 0.03  & 0.08  & 0.16  & 0.17  \\ 
PID efficiencies                         & 0.03  & 0.03  & 0.01  & 0.05  & 0.02  & 0.02  \\ 
Fixed yield ratios                       & 0.05  & 0.06  & 0.03  & 0.06  & 0.02  & 0.02  \\ 
Mass-shape bin dependence                & 0.05  & 0.07  & 0.04  & 0.08  & 0.07  & 0.09  \\ 
Part. reco. physics effects              & 0.04  & 0.10  & 0.15  & 0.05  & 0.10  & 0.09  \\ 
Small backgrounds                        & 0.11  & 0.16  & 0.13  & 0.12  & 0.08  & 0.13  \\ 
Dalitz-bin migration                     & 0.04  & 0.08  & 0.08  & 0.11  & 0.18  & 0.10  \\ 
\CP violation of \KS                     & 0.03  & 0.04  & 0.08  & 0.08  & 0.09  & 0.46  \\ 
\D mixing                                & 0.04  & 0.01  & 0.00  & 0.02  & 0.02  & 0.01  \\ 
Bias correction                          & 0.04  & 0.03  & 0.02  & 0.04  & 0.09  & 0.05  \\ 

\midrule
Total LHCb-related uncertainty                   & 0.20  & 0.25  & 0.24  & 0.26  & 0.32  & 0.54  \\ 
\midrule
Total systematic uncertainty                        & 0.31  & 0.43  & 0.30  & 0.38  & 0.35  & 0.57  \\ 

\bottomrule
\end{tabular}

\end{table}

Systematic uncertainties on the measurements of the \CP observables are evaluated and are presented in Table~\ref{tab:systematic_uncertainties}. 
The limited precision on (\ci, \si) coming from the combined \besiii and \cleo~\cite{bes3prd,Krishna} results induces uncertainties on the \CP parameters. These uncertainties are evaluated by fitting the data multiple times, each time with different (\ci, \si) values sampled according to their experimental uncertainties and correlations.\footnote{The detailed output of this study is available as supplementary material to this paper at the publisher's \href{https://link.springer.com/article/10.1007/JHEP02(2021)169}{\texttt{website}}, and provides sufficient information to determine the correlation between this uncertainty and the corresponding uncertainties of future \g measurements that also rely on the same strong-phase measurements.
}
The resulting standard deviation of each distribution of the \CP observables is assigned as the systematic uncertainty. The size of the systematic uncertainty is notably much smaller than the corresponding uncertainty in Ref.~\cite{LHCb-PAPER-2018-017} due to the improvement in the knowledge of these strong-phase parameters~\cite{bes3prd, Krishna}. 

The non-uniform efficiency profile over the Dalitz plot means that the values of (\ci,\si) appropriate for this analysis can differ from those measured in Refs.~\cite{bes3prd,Krishna}, which correspond to the case where there is no variation in efficiency over the Dalitz plot. Amplitude models from Refs.~\cite{Belle2018,BABAR2010} are used to calculate the values of \ci and \si both with and without the efficiency profiles determined from simulation. The shift in the \ci and \si values is taken as an estimate of the size of this effect. Pseudoexperiments are generated assuming the shifted \ci and \si values and fit with the default values of \ci and \si. The mean bias of each \CP observable is assigned as a systematic uncertainty. 
The assumption that the relative variation of efficiency over the Dalitz plot is the same in selected \BtoDK and \BtoDpi candidates is verified in simulated samples of similar size to the \BtoDpi yields observed in data. No statistically significant difference is observed and no systematic uncertainty is assigned. 

The uncertainties from the fixed invariant-mass shapes determined in the global fit are propagated to the \CP observables through a resampling method~\cite{ChernickMichaelR1999Bm:a}. The following procedure, which takes into account the fact that some parameters are determined in simulation and others in data, is carried out a hundred times. First, the simulated decays that were used to determine the nominal mass shape parameters are each resampled with replacement and fit to determine an alternative set of parameters. Then, the final dataset is resampled with replacement and the global fit is repeated using the alternative fixed shape parameters, to determine alternative values for the parameters that are determined from real data. Finally, the \CP fit is  performed using the alternative invariant-mass parameterisations, without resampling the final dataset. The standard deviation of the \CP observables obtained via this procedure is taken as the systematic uncertainty due to the fixed parameterisation. 

The PID efficiencies are varied within their uncertainties in the global and \CP fit and the standard deviation of the \CP parameters is taken as the systematic uncertainty. A similar method is used to determine the uncertainties due to the fixed fractions between different partially reconstructed backgrounds where the uncertainties on the fixed fractions are those from the branching fractions~\cite{PDG2020} and the selection efficiencies. 

The \CP fit assumes the same mass shape for each component in each Dalitz plot bin. For the signal and cross-feed backgrounds the shapes are redetermined in each bin using the same procedures described in Sect.~\ref{sec:massfit}. The variance is very small due to weak correlations between phase-space coordinates and particle kinematics. The combinatorial slope can also vary from bin to bin, as the relative rate of combinatorial background with and without a real \Dz meson will not be constant. The size of this effect is determined through the study of the high invariant-mass sideband where only combinatorial background contributes. Pseudodata are generated where this variation in mass shape across the Dalitz plot bins is replicated for signal, cross-feed and combinatorial backgrounds, and the generated samples are fit with the default fit assumptions of the same shape in each bin. The mean bias is assigned as the systematic uncertainty.

The partially reconstructed background shape is also expected to vary in each bin, however the leading source of this effect is due to the individual components of this background having a different distribution over the Dalitz plot. Some partially reconstructed backgrounds will be distributed as \Dz(\Dzb) \to \Kshh for reconstructed \Bm (\Bp) candidates, while others will be distributed as a $\Dz$--$\Dzb$ admixture depending on the relevant \CP-violation parameters. Pseudodata are generated, where the $D$-decay phase-space distributions for $\Bpm\to \D^*K^\pm$ and $\Bpm \to \D K^{*+}$ background events are based on the \CP parameters reported in Ref.~\cite{LHCb-CONF-2018-002}. No \CP violation is introduced into the partially reconstructed background in the \BtoDpi samples since it is expected to be small, and the $\Bz \to \D \rho^0$ background is treated as an equal mix of \Dz and \Dzb since either pion can be reconstructed. The generated pseudodata are fitted with the default fit and the mean bias is assigned as the systematic uncertainty. 

Systematic uncertainties are assigned for small residual backgrounds that contaminate the data sample but are not accounted for in the fit. Their impact is assessed by generating pseudoexperiments that contain these backgrounds and are fit with the default model. The mean bias is assigned as the uncertainty. One source of background is from $\Lb \to \D \Pp \pim$ decays where the pion is not reconstructed and the proton is misidentified as a kaon. This background is modelled as a \Dzb-like contribution in \Bm decays, and has an expected yield of 0.5\% of the \BtoDK signal. A further, even smaller, background is $\Lb \to \Lc (\to p \KS \pip\pim)\pim$ decays  where the \pip meson in the $\Lambda_c^+$ decay is missed, and the \Pp reconstructed as the \pip from the \D-decay. The effective distribution of the reconstructed \D meson is unknown and is assigned to be \Dzb-like in \Bm decays to be conservative. The mass shapes and rates of these backgrounds are determined from simulation. Another source of background comes from residual $B\to\D\mu\nu$ decays, where the rate (less than 0.2\,\% relative to the signal mode, after the applied veto) and shape are determined from simulation with PID efficiencies from calibration data. The residual semileptonic $\D$ decay background has a rate of less than 0.1\% of signal and the distribution of these events on the Dalitz plot is determined through a simplified simulation~\cite{Cowan:2016tnm} taking into account various $K^{*}$ mesons. Finally, a small peaking background from $\Bpm \to \D(\to \Kpm\pimp)\KS\pipm$ decays where the
kaon is reconstructed as the companion and the other particles are assigned to the \D decay is considered. The yield of this background is determined to be 0.5\% of the signal yield in \BtoDK by a data driven study of the invariant-mass distribution of switched tracks. The distribution on the Dalitz plot is determined through the simplified simulation~\cite{Cowan:2016tnm} where different $K^{*\pm}\to \KS\pipm$ resonances are generated. 

The main effect of migration from one Dalitz plot bin to another is implicitly taken into account by using the data to determine the $F_i$, which thus include the effects of the net bin migration. However, a small effect arises because of the differences in the distributions of the \BtoDK and \BtoDpi decays due to the differing hadronic decay parameters. To investigate this, data points are generated according to the amplitude model in Ref.~\cite{Belle2018} with \CP observables consistent with expectation~\cite{LHCb-CONF-2018-002,HFLAV18}. To smear these data points on the Dalitz plot, an event is selected from full LHCb simulation and the difference in \msqplus and \msqmin between its true and reconstructed quantities is applied to the data point in order to determine its reconstructed bin. The difference between true and reconstructed quantities is multiplied by a factor of 1.2 to account for differences in resolution between data and simulation. Pseudoexperiments are generated based on the expected reconstructed yields in each bin and fit with a nominal fit where the \ci and \si parameters are determined by the amplitude model~\cite{Belle2018}. The mean bias in the \CP violation parameters is taken as the systematic uncertainty, which is small. 

The impact of ignoring the \CP violation and matter effects in \KS decays is determined through generating pseudoexperiments taking into account all these effects as detailed in Ref.~\cite{KsCPV}, where LHCb simulation is used to obtain the \KS lifetime acceptance and momentum distribution. The size of the bias found is consistent with those expected from Ref.~\cite{KsCPV}, where it was also predicted that the relative uncertainties on \BtoDpi observables are be expected to be larger than for \BtoDK observables. This is found to be true, but even the most significant uncertainty, on \yxidpi, is an order of magnitude smaller than the corresponding statistical uncertainty. The effect of ignoring charm mixing is expected to be minimal, given that the first-order effects are inherently taken into account when the \Fi parameters are measured as a part of the fit~\cite{BPV}. This is verified by generating pseudoexperiments that include charm mixing and fitting them with the nominal fit. 

In previous studies, a bias correction has been necessary when similar measurements have been performed with lower signal yields~\cite{LHCb-PAPER-2018-017} leading to some fit instabilities. In this case, the higher yields have resulted in a bias that is of negligible size and hence no correction is applied. Nonetheless, the uncertainty on the biases are assigned as the systematic uncertainties. 

In general, all the systematic uncertainties are small in comparison to the statistical uncertainties. There is no dominant source of systematic uncertainty for all \CP observables, however the description of backgrounds, either those not modelled or the modelling of the partially reconstructed backgrounds are some of the larger sources. The uncertainty attributed to the precision of the strong-phase measurements is of similar size to the total LHCb-related systematic uncertainty.

\section{Interpretation}
\label{sec:interpretation}

The \CP observables are measured to be
\textbf{}\begin{align}
\begin{split}
    x_-^{DK} & = (\phantom{-}5.68 \pm 0.96 \pm  0.20\pm 0.23) \times 10^{-2}, \\
    y_-^{DK} & = (\phantom{-}6.55 \pm 1.14 \pm  0.25\pm 0.35) \times 10^{-2}, \\
    x_+^{DK} & = (         - 9.30 \pm 0.98 \pm  0.24\pm 0.18) \times 10^{-2}, \\
    y_+^{DK} & = (         - 1.25 \pm 1.23 \pm  0.26\pm 0.28) \times 10^{-2}, \\
    x_\xi^{D\pi} & = (         - 5.47 \pm 1.99 \pm  0.32\pm 0.14) \times 10^{-2}, \\
    y_\xi^{D\pi} & = (\phantom{-}0.71 \pm 2.33 \pm  0.54\pm 0.18) \times 10^{-2}, 
\end{split}
\end{align}
where the first uncertainty is statistical, the second arises from systematic effects in the method or detector considerations, and the third from external inputs of strong-phase measurements from the combination of \cleo and \besiii~\cite{bes3prd,CLEOCISI} results. The correlation matrices for each source of uncertainty are available in the appendices in Tables~\ref{tab:stat_corr_matrix}-\ref{tab:cisi_syst_corr_matrix}. 

\begin{figure}
    \centering
    \includegraphics[width=0.7\columnwidth]{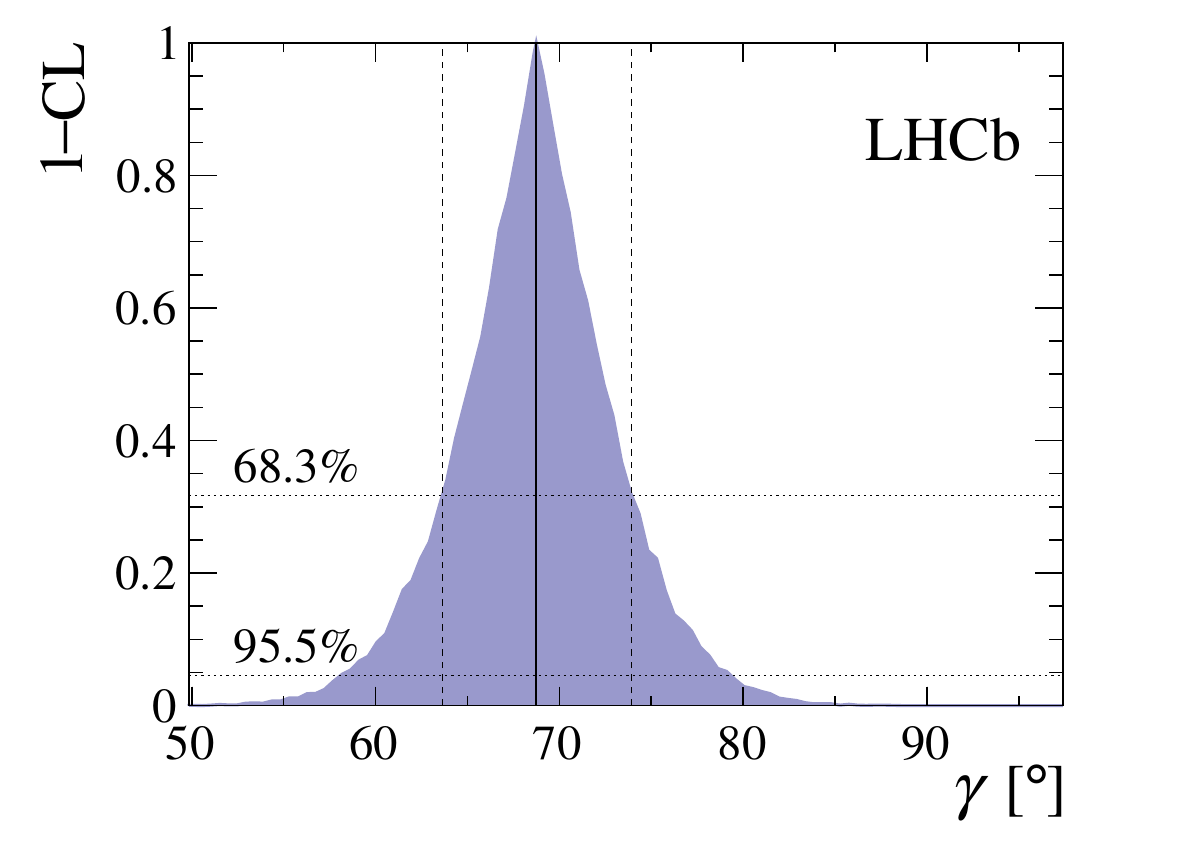}

    \caption{Confidence limits for the CKM angle $\gamma$ obtained using the method described in Ref.~\cite{LHCb-PAPER-2016-032}.}
    \label{fig:interpretation_1d}
\end{figure}

\begin{figure}
    \centering
    \includegraphics[width=0.45\columnwidth]{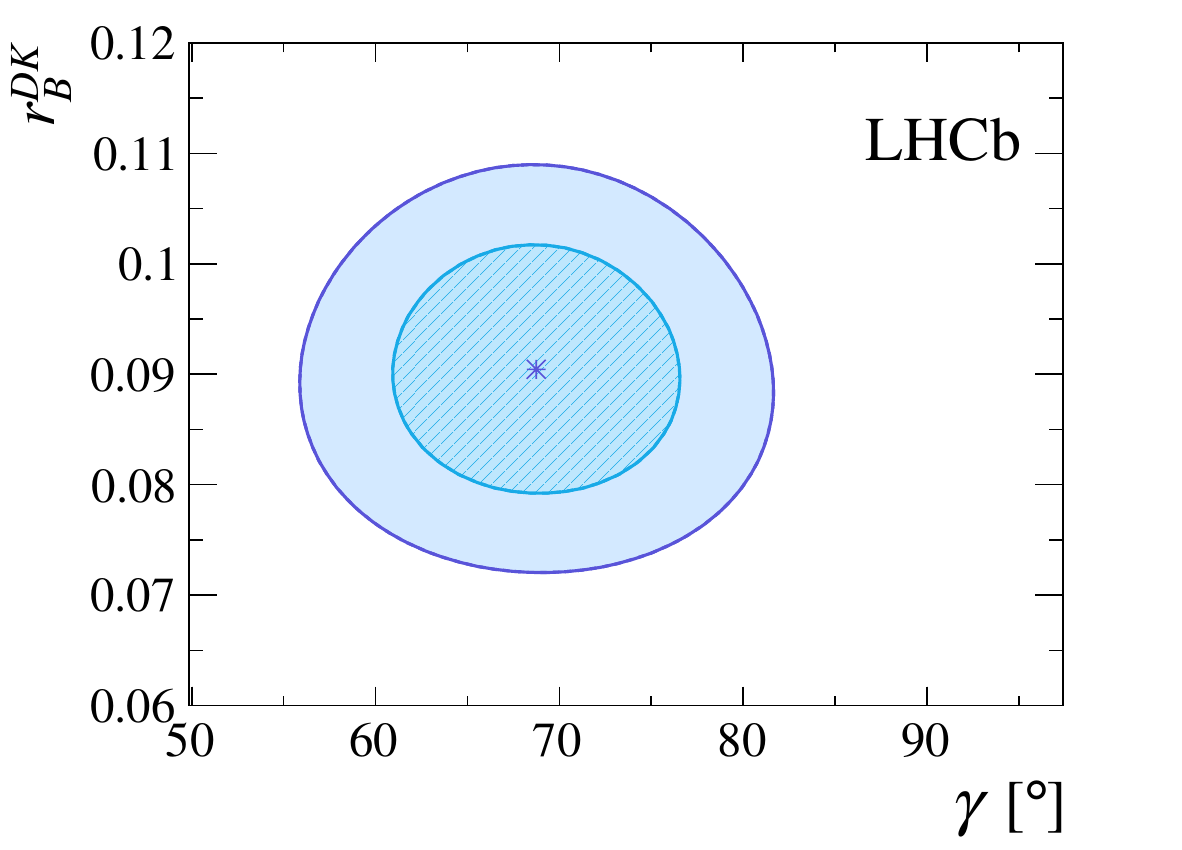}
    \includegraphics[width=0.45\columnwidth]{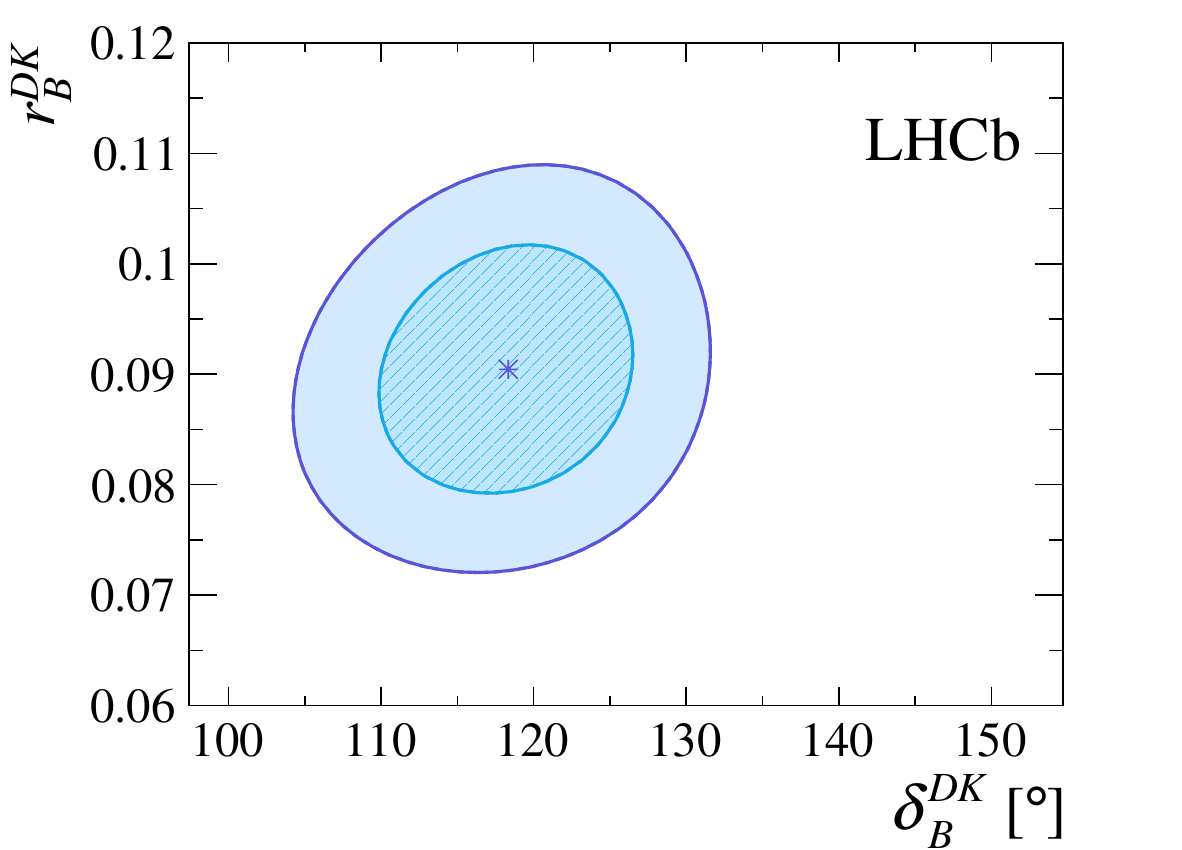}
    \includegraphics[width=0.45\columnwidth]{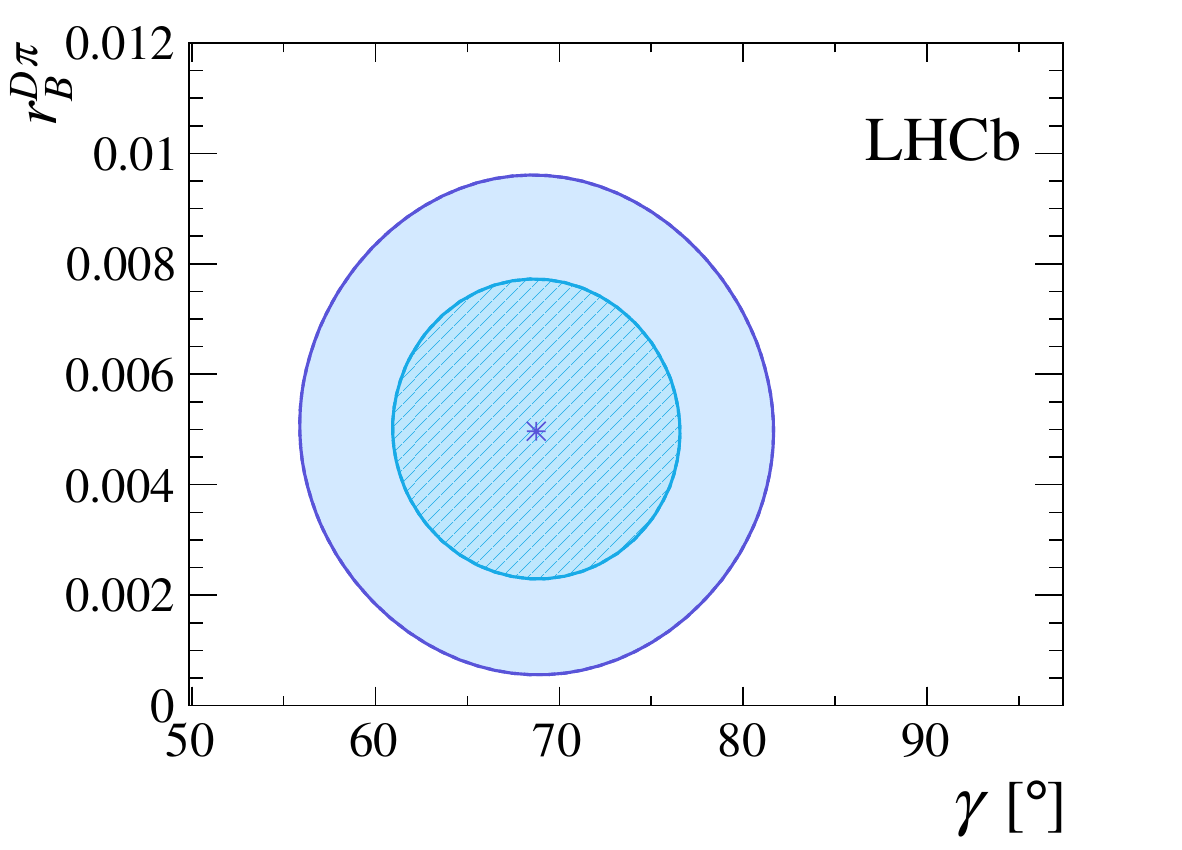}
    \includegraphics[width=0.45\columnwidth]{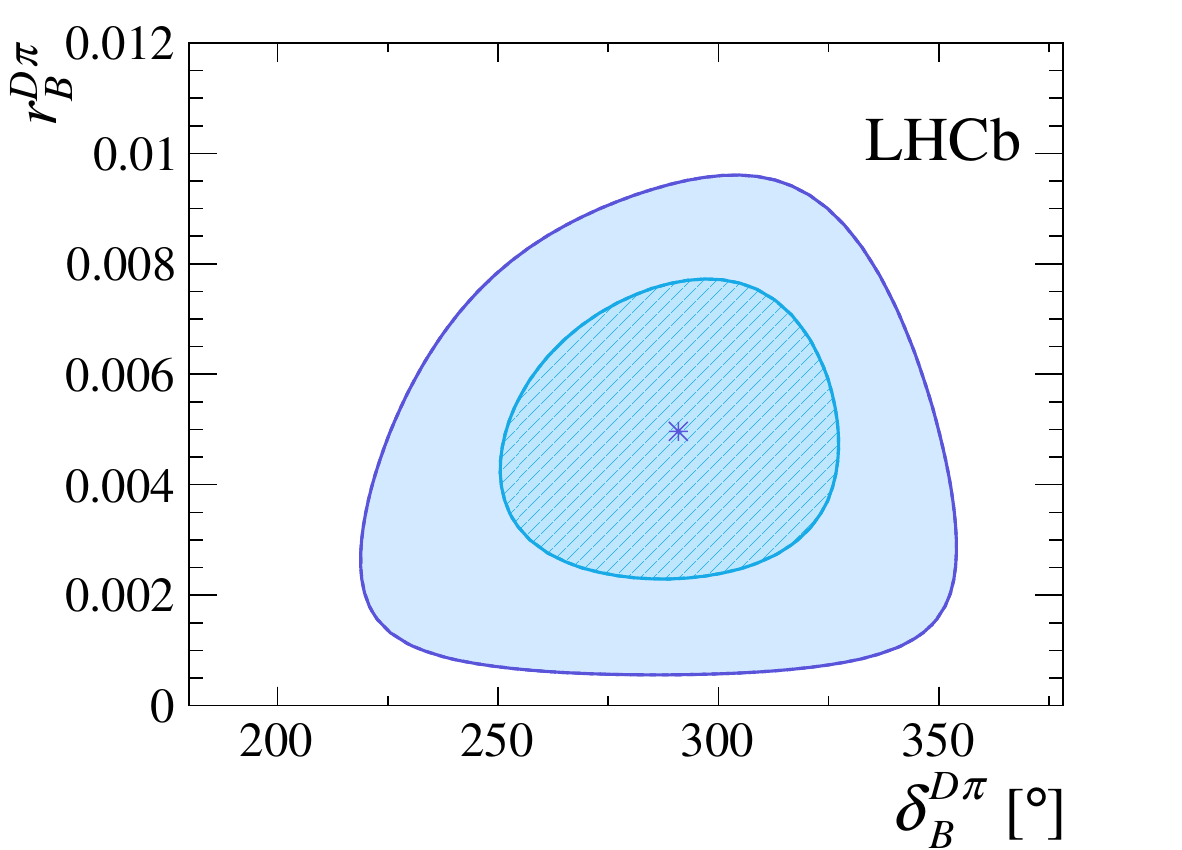}
    \caption{The 68\,\% and 95\,\% confidence regions for combinations of the physics parameters $(\g, \rBDK, \dBDK,\rBDpi, \dBDpi)$ obtained using the methods described in Ref.~\cite{LHCb-PAPER-2016-032}.}
    \label{fig:interpretation}
\end{figure}

The \CP observables are interpreted in terms of the underlying physics parameters \g, and \rB and \dB for each \Bpm decay mode. The interpretation is done via a maximum likelihood fit using a frequentist treatment as described in Ref.~\cite{LHCb-PAPER-2016-032}. The solution for the physics parameters has a two-fold ambiguity as the equations are invariant under the simultaneous substitutions $\g \to \g+180^\circ$ and $\dB \to \dB + 180^\circ$. The solution that satisfies $0 < \g < 180^\circ$ is chosen, and leads to

\begin{align}
 \begin{split}\label{eq:phys_results}
     \gamma          &= (68.7^{+5.2}_{-5.1})^\circ, \\
     r_B^{\DK}       &= 0.0904^{+0.0077}_{-0.0075}, \\
     \delta_B^{\DK}  &= (118.3^{+5.5}_{-5.6})^\circ, \\
     r_B^{\Dpi}      &= 0.0050 \pm 0.0017, \\
     \delta_B^{\Dpi} &= (291^{+24}_{-26})^\circ.
 \end{split}
 \end{align}
Pseudoexperiments are carried out to confirm that the value of \Pgamma is extracted without bias.  
This is the most precise single measurement of $\gamma$ to date. The result is consistent with the indirect determination $\g= \left(65.66^{+0.90}_{-2.65}\right)^\circ$~\cite{CKMfitter2005}. The confidence limits for $\gamma$ are illustrated in Fig.~\ref{fig:interpretation_1d}, while Fig.~\ref{fig:interpretation} shows the two-dimensional confidence regions obtained for the  (\g, \rB) and (\rB, \dB) parameter combinations. The results for \g, $\rBDK$ ,and $\dBDK$ are consistent with their current world averages~\cite{HFLAV18,CKMfitter2005} which include the \lhcb results obtained with the 2011--2016 data. The knowledge of $\rBDpi$ and $\dBDpi$ from other sources is limited, with the combination of many observables presented in Ref.~\cite{LHCb-PAPER-2016-032} providing two possible solutions. The results here have a single solution, and favour a central value that is consistent with the expectation for $\rBDpi$, given the value of $\rBDK$ and CKM elements~\cite{Kenzie:2016yee}.  This is likely to remove the two-solution aspect in future combinations of \g and associated hadronic parameters. The low value of \rBDpi means that the direct contribution to \g from \BtoDpi decays in this measurement is minimal. However the ability to use this decay mode to determine the efficiency has approximately halved the total LHCb related experimental systematic uncertainty in comparison to Ref.~\cite{LHCb-PAPER-2018-017}. The new inputs from the \besiii collaboration have led to the strong-phase related uncertainty on \g to be approximately 1$^\circ$, which is a significant reduction compared to the propagated uncertainty when only \cleo measurements were available.  

\section{Conclusions}
\label{sec:conclusions}

In summary, the decays  \BtoDK and \BtoDpi  with \DtoKspp or ${\DtoKsKK}$ obtained from the full \lhcb dataset collected to date, corresponding to an integrated luminosity of 9~\invfb, have been analysed to determine the CKM angle \g. The sensitivity to \g comes almost entirely from \BtoDK decays where the signal yields of reconstructed events are approximately 13600 (1900) in the \DtoKspp (\DtoKskk) decay modes. The \BtoDpi data is primarily used to control effects due to selection and reconstruction of the data, which leads to small experimental systematic uncertainties. The analysis is performed in bins of the \D-decay Dalitz plot and a combination of measurements performed by the \cleo and \besiii collaborations presented in Refs.~\cite{bes3prd,Krishna} are used to provide input on the \D-decay strong-phase parameters (\ci,\si). Such an approach allows the analysis to be free from model-dependent assumptions on the strong-phase variation across the Dalitz plot. The analysis also determines the hadronic parameters \rB and \dB for each \Bpm decay mode. Those of the \BtoDK decay are consistent with current averages, and those of the \BtoDpi decay are obtained with the best precision to date, and have not previously been measured using these \D-decay modes. The CKM angle \g is determined to be $\g = (68.7^{+5.2}_{-5.1})^\circ$, where the result is limited by statistical uncertainties. This is the most precise measurement of \g from a single analysis, and supersedes the results in Refs.~\cite{LHCb-PAPER-2018-017, LHCb-PAPER-2014-041}.

\section*{Acknowledgements}
%
%
\noindent We express our gratitude to our colleagues in the CERN
accelerator departments for the excellent performance of the LHC. We
thank the technical and administrative staff at the LHCb
institutes.
We acknowledge support from CERN and from the national agencies:
CAPES, CNPq, FAPERJ and FINEP (Brazil); 
MOST and NSFC (China); 
CNRS/IN2P3 (France); 
BMBF, DFG and MPG (Germany); 
INFN (Italy); 
NWO (Netherlands); 
MNiSW and NCN (Poland); 
MEN/IFA (Romania); 
MSHE (Russia); 
MICINN (Spain); 
SNSF and SER (Switzerland); 
NASU (Ukraine); 
STFC (United Kingdom); 
DOE NP and NSF (USA).
We acknowledge the computing resources that are provided by CERN, IN2P3
(France), KIT and DESY (Germany), INFN (Italy), SURF (Netherlands),
PIC (Spain), GridPP (United Kingdom), RRCKI and Yandex
LLC (Russia), CSCS (Switzerland), IFIN-HH (Romania), CBPF (Brazil),
PL-GRID (Poland) and OSC (USA).
We are indebted to the communities behind the multiple open-source
software packages on which we depend.
Individual groups or members have received support from
AvH Foundation (Germany);
EPLANET, Marie Sk\l{}odowska-Curie Actions and ERC (European Union);
A*MIDEX, ANR, Labex P2IO and OCEVU, and R\'{e}gion Auvergne-Rh\^{o}ne-Alpes (France);
Key Research Program of Frontier Sciences of CAS, CAS PIFI,
Thousand Talents Program, and Sci. \& Tech. Program of Guangzhou (China);
RFBR, RSF and Yandex LLC (Russia);
GVA, XuntaGal and GENCAT (Spain);
the Royal Society
and the Leverhulme Trust (United Kingdom).

\clearpage
\section*{Appendices}

\appendix
\section{Correlation matrices}

The correlations matrices for the measured observables are shown in Tables~\ref{tab:stat_corr_matrix}--\ref{tab:cisi_syst_corr_matrix} for the statistical uncertainties, the experimental systematic uncertainties, and the strong-phase-related uncertainties, respectively.

\begin{table}[p]
\centering
\caption{Statistical uncertainties, $\sigma$, and correlation matrix for \xpmdk, \ypmdk, \xxidpi, and \yxidpi. 
\label{tab:stat_corr_matrix}}

\begin{tabular}{l|cccccc}
\toprule
\multicolumn{7}{c}{Uncertainty $(\times 10^{-2})$}\\ \midrule
 & \xmdk & \ymdk & \xpdk & \ypdk & \xxidpi & \yxidpi \\
\midrule
$\sigma$ & $0.96$ & $1.14$ & $0.98$ & $1.23$ & $1.99$ & $2.33$ \\
\bottomrule \multicolumn{7}{c}{} \\
\multicolumn{7}{c}{Correlation matrix}\\ \midrule
  & \xmdk & \ymdk & \xpdk & \ypdk & \xxidpi & \yxidpi \\
\midrule
$\xmdk$    & $\phantom{-}1\phantom{.000}$ & $-0.125$ & $-0.013$ & $\phantom{-}0.019$ & $\phantom{-}0.037$ & $-0.161$ \\[4pt]
$\ymdk $    &    & $\phantom{-}1\phantom{.000}$ & $-0.011$ & $-0.010$ & $\phantom{-}0.097$ & $\phantom{-}0.041$ \\[4pt]
$\xpdk $    & &       & $\phantom{-}1\phantom{.000}$ & $\phantom{-}0.105$ & $-0.108$ & $\phantom{-}0.032$ \\[4pt]
$\ypdk $    & & &          & $\phantom{-}1\phantom{.000}$ & $-0.070$ & $-0.147$ \\[4pt]
$\xxidpi$ & & &     &         & $\phantom{-}1\phantom{.000}$ & $\phantom{-}0.150$ \\[4pt]
$\yxidpi$ & & &     &    &         & $\phantom{-}1\phantom{.000}$ \\
\bottomrule
\end{tabular}

\end{table}

\begin{table}[tb]
\centering
\caption{Total \lhcb-related systematic uncertainties, $\sigma$, for \xpmdk, \ypmdk, \xxidpi, and \yxidpi, and the corresponding correlation matrix. 
\label{tab:lhcb_syst_corr_matrix}}

      \begin{tabular}{l|cccccc}
    \toprule
    \multicolumn{7}{c}{Uncertainty $(\times 10^{-2})$}\\ \midrule
     & \xmdk & \ymdk & \xpdk & \ypdk & \xxidpi & \yxidpi \\
    \midrule
    $\sigma$ & $0.20$ & $0.25$ & $0.24$ & $0.26$ & $0.32$ & $0.54$ \\
    \bottomrule \multicolumn{7}{c}{} \\
    \multicolumn{7}{c}{Correlation matrix}\\ \midrule
     & \xmdk & \ymdk & \xpdk & \ypdk & \xxidpi & \yxidpi \\
    \midrule
    $\xmdk$    & $\phantom{-}1\phantom{.000}$ & $\phantom{-}0.864$ & $\phantom{-}0.734$ & $\phantom{-}0.897$ & $\phantom{-}0.349$ & $\phantom{-}0.318$ \\[4pt]
    $\ymdk $    &    & $\phantom{-}1\phantom{.000}$ & $\phantom{-}0.874$ & $\phantom{-}0.903$ & $\phantom{-}0.408$ & $\phantom{-}0.362$ \\[4pt]
    $\xpdk$    & &       & $\phantom{-}1\phantom{.000}$ & $\phantom{-}0.771$ & $\phantom{-}0.563$ & $\phantom{-}0.447$ \\[4pt]
    $\ypdk$    & & &          & $\phantom{-}1\phantom{.000}$ & $\phantom{-}0.507$ & $\phantom{-}0.451$ \\[4pt]
    $\xxidpi$ & & &     &         & $\phantom{-}1\phantom{.000}$ & $\phantom{-}0.484$ \\[4pt]
    \yxidpi & & &     &    &         & $\phantom{-}1\phantom{.000}$ \\
    \bottomrule
    \end{tabular}

\end{table}

\begin{table}[tb]
\centering
\caption{Systematic uncertainties, $\sigma$, for \xpmdk, \ypmdk, \xxidpi, and \yxidpi  due to strong-phase inputs, the corresponding correlation matrix. 
\label{tab:cisi_syst_corr_matrix}}

\begin{tabular}{l|cccccc}
\toprule
\multicolumn{7}{c}{Uncertainty $(\times 10^{-2})$}\\ \midrule
  & \xmdk & \ymdk & \xpdk & \ypdk & \xxidpi & \yxidpi \\
\midrule
$\sigma$ & $0.23$ & $0.35$ & $0.18$ & $0.28$ & $0.14$ & $0.18$ \\
\bottomrule \multicolumn{7}{c}{} \\
\multicolumn{7}{c}{Correlation matrix}\\ \midrule
  & \xmdk & \ymdk & \xpdk & \ypdk & \xxidpi & \yxidpi \\
\midrule
$\xmdk $    & $\phantom{-}1\phantom{.000}$ & $-0.047$ & $-0.490$ & $\phantom{-}0.322$ & $\phantom{-}0.189$ & $\phantom{-}0.144$ \\ [4pt]
$\ymdk$    &    & $\phantom{-}1\phantom{.000}$ & $\phantom{-}0.059$ & $-0.237$ & $-0.116$ & $-0.117$ \\ [4pt]
$\xpdk $    & &       & $\phantom{-}1\phantom{.000}$ & $\phantom{-}0.061$ & $\phantom{-}0.004$ & $-0.139$ \\ [4pt]
$\ypdk$    & & &          & $\phantom{-}1\phantom{.000}$ & $\phantom{-}0.127$ & $-0.199$ \\ [4pt]
$\xxidpi$ & & &     &         & $\phantom{-}1\phantom{.000}$ & $\phantom{-}0.638$ \\[4pt]
$\yxidpi$ & & &     &    &         & $\phantom{-}1\phantom{.000}$ \\
\bottomrule
\end{tabular} 

\end{table}

\clearpage

\addcontentsline{toc}{section}{References}
\bibliographystyle{LHCb}
\bibliography{main,standard,LHCb-PAPER,LHCb-CONF,LHCb-DP,LHCb-TDR}

\newpage
\centerline
{\large\bf LHCb collaboration}
\begin
{flushleft}
\small
R.~Aaij$^{31}$,
C.~Abell{\'a}n~Beteta$^{49}$,
T.~Ackernley$^{59}$,
B.~Adeva$^{45}$,
M.~Adinolfi$^{53}$,
H.~Afsharnia$^{9}$,
C.A.~Aidala$^{84}$,
S.~Aiola$^{25}$,
Z.~Ajaltouni$^{9}$,
S.~Akar$^{64}$
% ...

{\footnotesize \it

% ...
$ ^{9}$Universit{\'e} Clermont Auvergne, CNRS/IN2P3, LPC, Clermont-Ferrand, France\\
% ...
$ ^{25}$INFN Sezione di Milano, Milano, Italy\\
% ...
$ ^{31}$Nikhef National Institute for Subatomic Physics, Amsterdam, Netherlands\\
% ...
$ ^{45}$Instituto Galego de F{\'\i}sica de Altas Enerx{\'\i}as (IGFAE), Universidade de Santiago de Compostela, Santiago de Compostela, Spain\\
% ...
$ ^{49}$Physik-Institut, Universit{\"a}t Z{\"u}rich, Z{\"u}rich, Switzerland\\
% ...
$ ^{53}$H.H. Wills Physics Laboratory, University of Bristol, Bristol, United Kingdom\\
% ...
$ ^{59}$Oliver Lodge Laboratory, University of Liverpool, Liverpool, United Kingdom\\
% ...
$ ^{64}$University of Cincinnati, Cincinnati, OH, United States\\
% ...
$ ^{67}$Syracuse University, Syracuse, NY, United States\\
% ...
$ ^{84}$University of Michigan, Ann Arbor, United States, associated to $^{67}$\\
\bigskip
% ...
\medskip
}
\end{flushleft}

\end{document}